%Paper: hep-th/9503027
%From: "Clifford Johnson" <cvj@puhep1.Princeton.EDU>
%Date: Sun, 5 Mar 95 17:54:35 -0500

%% Remove the following line if you cannot (or do not want to) include
%% the figure:
\input epsf
%% You should also remove the two occurences in the text of
%%`\epsfxsize=5.0in\epsfbox{throat.eps}' if not using epsf.tex

%%%%%%%%%%%%%%%%  preprint and letter macro  %%%%%%%%%%%%%%%%%

\hsize=6.0truein
\vsize=8.5truein
\voffset=0.25truein
\hoffset=0.1875truein%would be 0.25 except our laser printer is off by 1/16in
\tolerance=1000
\hyphenpenalty=500
\def\monthintext{\ifcase\month\or January\or February\or
   March\or April\or May\or June\or July\or August\or
   September\or October\or November\or December\fi}

%%%%%%%%%%%%%%%%%  Twelve point text font  %%%%%%%%%%%%%%%%%%%

\font\tenrm=cmr10 scaled \magstep1   \font\tenbf=cmbx10 scaled \magstep1
\font\sevenrm=cmr7 scaled \magstep1  
\font\fiverm=cmr5 scaled \magstep1   

\font\teni=cmmi10 scaled \magstep1   \font\tensy=cmsy10 scaled \magstep1
\font\seveni=cmmi7 scaled \magstep1  \font\sevensy=cmsy7 scaled \magstep1
\font\fivei=cmmi5 scaled \magstep1   \font\fivesy=cmsy5 scaled \magstep1

\font\tentt=cmtt10 scaled \magstep1
\font\tenit=cmti10 scaled \magstep1
\font\tensl=cmsl10 scaled \magstep1

\def\twelvepoint{\def\rm{\fam0\tenrm}
   \textfont0=\tenrm \scriptfont0=\sevenrm \scriptscriptfont0=\fiverm
   \textfont1=\teni  \scriptfont1=\seveni  \scriptscriptfont1=\fivei
   \textfont2=\tensy \scriptfont2=\sevensy \scriptscriptfont2=\fivesy
   \textfont\itfam=\tenit \def\it{\fam\itfam\tenit}
   \textfont\ttfam=\tentt \def\tt{\fam\ttfam\tentt}
   \textfont\bffam=\tenbf \def\bf{\fam\bffam\tenbf}
   \textfont\slfam=\tensl \def\sl{\fam\slfam\tensl} \rm
   %Essentially I changed all dimensions to 1.2 times as large as in plain tex
   \hfuzz=1pt\vfuzz=1pt%much more than plain tex's value
   \setbox\strutbox=\hbox{\vrule height 10.2pt depth 4.2pt width 0pt}
   \parindent=24pt\parskip=1.2pt plus 1.2pt
   \topskip=12pt\maxdepth=4.8pt\jot=3.6pt
   \normalbaselineskip=14.4pt\normallineskip=1.2pt
   \normallineskiplimit=0pt\normalbaselines
   \abovedisplayskip=13pt plus 3.6pt minus 5.8pt
   \belowdisplayskip=13pt plus 3.6pt minus 5.8pt
   \abovedisplayshortskip=-1.4pt plus 3.6pt
   \belowdisplayshortskip=13pt plus 3.6pt minus 3.6pt
   %plain tex's value for belowdisplayshortskip looked terrible
   \topskip=12pt \splittopskip=12pt
   \scriptspace=0.6pt\nulldelimiterspace=1.44pt\delimitershortfall=6pt
   \thinmuskip=3.6mu\medmuskip=3.6mu plus 1.2mu minus 1.2mu
   \thickmuskip=4mu plus 2mu minus 1mu%reduced these plain tex values
   \smallskipamount=3.6pt plus 1.2pt minus 1.2pt
   \medskipamount=7.2pt plus 2.4pt minus 2.4pt
   \bigskipamount=14.4pt plus 4.8pt minus 4.8pt}

\twelvepoint

%%%%%%%%%%%%%%%%% Definitions for Preprints %%%%%%%%%%%%%%%%%%

% title page title font

\font\titlerm=cmr10 scaled \magstep3
\font\titlerms=cmr10 scaled \magstep1 %\font\titlermss=cmr8
\font\titlei=cmmi10 scaled \magstep3  %math italic for title
\font\titleis=cmmi10 scaled \magstep1 %\font\titleiss=cmmi8
\font\titlesy=cmsy10 scaled \magstep3 	%math symbols for title
\font\titlesys=cmsy10 scaled \magstep1  %\font\titlesyss=cmsy8
\font\titleit=cmti10 scaled \magstep3	%text italic for title
\skewchar\titlei='177 \skewchar\titleis='177 %\skewchar\titleiss='177
\skewchar\titlesy='60 \skewchar\titlesys='60 %\skewchar\titlesyss='60

\def\titlefont{\def\rm{\fam0\titlerm}% switch to title font
   \textfont0=\titlerm \scriptfont0=\titlerms %\scriptscriptfont0=\titlermss
   \textfont1=\titlei  \scriptfont1=\titleis  %\scriptscriptfont1=\titleiss
   \textfont2=\titlesy \scriptfont2=\titlesys %\scriptscriptfont2=\titlesyss
   \textfont\itfam=\titleit \def\it{\fam\itfam\titleit} \rm}

% title page macros

\def\preprint#1{\baselineskip=19pt plus 0.2pt minus 0.2pt \pageno=0
   \begingroup%use with \draft or \date to end group
   \nopagenumbers\parindent=0pt\baselineskip=14.4pt\rightline{#1}}
\def\title#1{
   \vskip 0.9in plus 0.45in
   \centerline{\titlefont #1}}
\def\secondtitle#1{}%set up this some time
\def\author#1#2#3{\vskip 0.9in plus 0.45in
   \centerline{{\bf #1}\myfoot{#2}{#3}}\vskip 0.12in plus 0.02in}
\def\secondauthor#1#2#3{}%set up this some time
\def\addressline#1{\centerline{#1}}
\def\abstract{\vskip 0.7in plus 0.35in
	\centerline{\bf Abstract}
	\smallskip}
\def\finishtitlepage#1{\vskip 0.8in plus 0.4in
   \leftline{#1}\supereject\endgroup}

\def\date#1{\finishtitlepage{#1}}

\def\nolabels{\def\eqnlabel##1{}\def\eqlabel##1{}\def\figlabel##1{}%
	\def\reflabel##1{}}
\def\writelabels{\def\eqnlabel##1{%
	{\escapechar=` \hfill\rlap{\hskip.11in\string##1}}}%
	\def\eqlabel##1{{\escapechar=` \rlap{\hskip.11in\string##1}}}%
	\def\figlabel##1{\noexpand\llap{\string\string\string##1\hskip.66in}}%
	\def\reflabel##1{\noexpand\llap{\string\string\string##1\hskip.37in}}}
\nolabels

%  tagged section numbers

\global\newcount\secno \global\secno=0
\global\newcount\meqno \global\meqno=1
\global\newcount\subsecno \global\subsecno=0
\global\newcount\subsubsecno \global\subsubsecno=0

\font\secfont=cmbx12 scaled\magstep1

\def\section#1{\global\advance\secno by1
   \xdef\secsym{\the\secno.}
   \global\subsecno=0
   \global\meqno=1\bigbreak\medskip
   \noindent{\secfont\the\secno. #1}\par\nobreak\smallskip\nobreak\noindent}
%\xdef\secsym{}

\def\subsection#1{\global\advance\subsecno by1
    %\xdef\secsym{\the\subsecno}
\global\subsubsecno=0
\medskip
\noindent
{\bf\the\secno.\the\subsecno\ #1}
\par\medskip\nobreak\noindent}
%\xdef\secsym{}

\def\subsubsection#1{\global\advance\subsubsecno by1
    %\xdef\secsym{\the\subsecno}
\medskip
\noindent{\sl\the\secno.\the\subsecno.\the\subsubsecno\ #1}
\par
\nobreak
\smallskip
\nobreak\noindent}

\def\acknowledgements{\bigbreak\medskip\noindent{\secfont
   Acknowledgements}\par\nobreak\smallskip\nobreak\noindent}

\def\newsec#1{\global\advance\secno by1
   \xdef\secsym{\the\secno.}
   \global\meqno=1\bigbreak\medskip
   \noindent{\bf\the\secno. #1}\par\nobreak\smallskip\nobreak\noindent}
\xdef\secsym{}

\def\appendix#1#2{\global\meqno=1\xdef\secsym{\hbox{#1.}}\bigbreak\medskip
\noindent{\bf Appendix #1. #2}\par\nobreak\smallskip\nobreak\noindent}

%\def\acknowledgements{\bigbreak\medskip\centerline{\bf
 %  Acknowledgements}\par\nobreak\smallskip\nobreak\noindent}

%         equations

\def\eqnn#1{\xdef #1{(\secsym\the\meqno)}%
	\global\advance\meqno by1\eqnlabel#1}
\def\eqna#1{\xdef #1##1{\hbox{$(\secsym\the\meqno##1)$}}%
	\global\advance\meqno by1\eqnlabel{#1$\{\}$}}
\def\eqn#1#2{\xdef #1{(\secsym\the\meqno)}\global\advance\meqno by1%
	$$#2\eqno#1\eqlabel#1$$}

%			 footnotes

\def\myfoot#1#2{{\baselineskip=14.4pt plus 0.3pt\footnote{#1}{#2}}}
%sequentially numbered footnotes
\global\newcount\ftno \global\ftno=1
\def\foot#1{{\baselineskip=14.4pt plus 0.3pt\footnote{$^{\the\ftno}$}{#1}}%
	\global\advance\ftno by1}

%         references

\global\newcount\refno \global\refno=1
\newwrite\rfile

\def\ref{[\the\refno]\nref}
\def\nref#1{\xdef#1{[\the\refno]}\ifnum\refno=1\immediate
	\openout\rfile=refs.tmp\fi\global\advance\refno by1\chardef\wfile=\rfile
	\immediate\write\rfile{\noexpand\item{#1\ }\reflabel{#1}\pctsign}\findarg}
%	horrible hack to sidestep tex \write limitation
\def\findarg#1#{\begingroup\obeylines\newlinechar=`\^^M\passarg}
	{\obeylines\gdef\passarg#1{\writeline\relax #1^^M\hbox{}^^M}%
	\gdef\writeline#1^^M{\expandafter\toks0\expandafter{\striprelax #1}%
	\edef\next{\the\toks0}\ifx\next\null\let\next=\endgroup\else\ifx\next\empty%

\else\immediate\write\wfile{\the\toks0}\fi\let\next=\writeline\fi\next\relax}}
	{\catcode`\%=12\xdef\pctsign{%}}
\def\striprelax#1{}

\def\semi{;\hfil\break}
\def\addref#1{\immediate\write\rfile{\noexpand\item{}#1}} %now unnecessary

\def\listrefs{\vfill\eject\immediate\closeout\rfile
   {{\secfont References}}\bigskip{\frenchspacing%
   \catcode`\@=11\escapechar=` %
   \input refs.tmp\vfill\eject}\nonfrenchspacing}

\def\startrefs#1{\immediate\openout\rfile=refs.tmp\refno=#1}

%		and finally, figures:

\global\newcount\figno \global\figno=1
\newwrite\ffile
\def\fig{\the\figno\nfig}
\def\nfig#1{\xdef#1{\the\figno}\ifnum\figno=1\immediate
	\openout\ffile=figs.tmp\fi\global\advance\figno by1\chardef\wfile=\ffile
	\immediate\write\ffile{\medskip\noexpand\item{Fig.\ #1:\ }%
	\figlabel{#1}\pctsign}\findarg}

\def\listfigs{\vfill\eject\immediate\closeout\ffile{\parindent48pt
	\baselineskip16.8pt{{\secfont Figure Captions}}\medskip
	\escapechar=` \input figs.tmp
%\vfill\eject
}}

%%%%%%%%%%%%%%%%%%%%%%%%%%%%%%%%%%%%%%%%%%%%%%%%%%%%%%%%%%%%%%%%%%%%%%%%%%%%%%%
\def\noblackbox{\overfullrule=0pt}
\def\inv{^{\raise.18ex\hbox{${\scriptscriptstyle -}$}\kern-.06em 1}}
\def\dup{^{\vphantom{1}}}
\def\Dsl{\,\raise.18ex\hbox{/}\mkern-16.2mu D} %this one can be subscripted
\def\dsl{\raise.18ex\hbox{/}\kern-.68em\partial}
\def\slash#1{\raise.18ex\hbox{/}\kern-.68em #1}
\def\lspace{}
\def\lbspace{}
\def\boxeqn#1{\vcenter{\vbox{\hrule\hbox{\vrule\kern3.6pt\vbox{\kern3.6pt
	\hbox{${\displaystyle #1}$}\kern3.6pt}\kern3.6pt\vrule}\hrule}}}
\def\mbox#1#2{\vcenter{\hrule \hbox{\vrule height#2.4in
	\kern#1.2in \vrule} \hrule}}  %e.g. \mbox{.1}{.1}
%matters of taste
%\def\tilde{\widetilde}
\def\bar{\overline}
\def\e#1{{\rm e}^{\textstyle#1}}
\def\del{\partial}
\def\curly#1{{\hbox{{$\cal #1$}}}}
\def\curlyD{\hbox{{$\cal D$}}}
\def\curlyL{\hbox{{$\cal L$}}}
\def\vev#1{\langle #1 \rangle}
\def\psibar{\overline\psi}
\def\lform{\hbox{$\sqcup$}\llap{\hbox{$\sqcap$}}}
\def\darr#1{\raise1.8ex\hbox{$\leftrightarrow$}\mkern-19.8mu #1}
\def\half{{\textstyle{1\over2}}} %puts a small half in a displayed eqn
\def\roughly#1{\ \lower1.5ex\hbox{$\sim$}\mkern-22.8mu #1\,}
\def\MSbar{$\bar{{\rm MS}}$}
%%%%%%%%%%%%%%%%%%%%%%%%%%%%%%%%%%%%%%%%%%%%%%%%%%%%%%%%%%%%%%
\hyphenation{di-men-sion di-men-sion-al di-men-sion-al-ly}
\noblackbox

\def\Tr{{\rm Tr}}
\def\det{{\rm det}}
\def\jump{\hskip1.0cm}
\def\wzw{Wess--Zumino--Witten}
\def\Az{A_z}
\def\Azb{A_{\bar{z}}}
\def\lr{\lambda_R}
\def\ll{\lambda_L}
\def\lrb{\bar{\lambda}_R}
\def\llb{\bar{\lambda}_L}
\font\top = cmbxti10 scaled \magstep1

\def\d{\partial_z}
\def\db{\partial_{\bar{z}}}
\def\rline{{{\rm I}\!{\rm R}}}
\def\tl{t_L}
\def\tr{t_R}
\def\eg{{\it e.g.,}\ }
\def\ie{{\it i.e.,}\ }
\def\cam{{\cal M}}
\def\cak{{\cal K}}

\preprint{
\vbox{\rightline{PUPT--1524}
\vskip2pt\rightline{McGill/95--01}
\vskip2pt\rightline{hep-th/9503027}
\vskip2pt\rightline{5th March 1995 %% Clifford's Birthday!!
}
}
}
\vskip -1.0cm
\title{A Conformal Field Theory of a Rotating  Dyon}

\vskip -1.5cm
\author{\bf Clifford V. Johnson\myfoot{$^*$}{\rm e-mail:
cvj@puhep1.princeton.edu}
and Robert C. Myers\myfoot{$^\dagger$}{\rm e-mail:
rcm@hep.physics.mcgill.ca}}{}{}
\vskip 0.3cm
\addressline{\it $^*$Joseph Henry Laboratories}
\addressline{\it Jadwin Hall}
\addressline{\it Princeton University}
\addressline{\it Princeton, NJ 08544 U.S.A.}
\bigskip
\addressline{\it $^{\dagger}$Department of Physics, McGill University}
\addressline{\it Ernest Rutherford Physics Building}
\addressline{\it 3600 University Street}
\addressline{\it Montr\'eal, Qu\'ebec, Canada H3A 2T8.}
\vskip -1.0cm
\abstract
\bigskip

A conformal field theory representing a
four--dimensional classical solution of
heterotic string theory is presented.
The low--energy limit of this solution
has  $U(1)$ electric and magnetic charges,
and also nontrivial axion and dilaton fields.
The low--energy metric contains mass, NUT and rotation parameters.
We demonstrate that this solution corresponds to
part of an  extremal  limit of the Kerr--Taub--NUT dyon
solution.  This limit displays interesting `remnant' behaviour,
in that  asymptotically far away from the dyon the
angular momentum vanishes, but far down the infinite throat in the
neighbourhood of the horizon (described by our CFT) there is a non--zero
angular velocity.
A further natural  generalization of the CFT
to include an additional
parameter is presented, but the full physical
 interpretation of its role in the
resulting low energy solution is unclear.
\vskip -1cm
\date{}
%\draft

\section{Introduction}

One of the important motivations to investigate string theory is the
expectation that it will provide a consistent quantum theory of gravity.
Thus the study of string propagation in curved space--times offers the
possibility that it may provide new insight into some of the longstanding
puzzles in quantum gravity, \eg the resolution of curvature
singularities, or a solution of
the information paradox in black hole thermodynamics.
In the context of string theory these topics are presently rich
sources of debate, conjecture and scientific inquiry\ref\debate{E. Witten,
{\sl `On black holes in string theory'}, Stony Brook, N.Y., Jun 1991.
published in Strings '91, Stony Brook, N.Y., 1991, hep-th/9111052\semi
T. Eguchi, {\sl Mod. Phys. Lett.} {\bf A7} (1992) 85, hep-th/9111001\semi
J. Ellis, N.E. Mavromatos and D.V. Nanopoulos, {\sl Phys. Lett.}
{\bf B284} (1992) 43, hep-th/9203012; {\sl Phys. Lett.} {\bf B276}
(1992) 56, hep-th/9111031\semi
L. Susskind, {\sl Phys. Rev. Lett.} {\bf 71} (1993) 2367, hep-th/9307168;
{\sl Phys. Rev.} {\bf D49} (1994) 6606, hep-th/9308139;
{\sl `Some speculations about black hole entropy in string theory'},
preprint RU-93-44, hep-th/9309145\semi
L. Susskind and J. Uglum, {\sl Phys. Rev.} {\bf D50} (1994) 2700,
hep-th/9401070\semi
E. Martinec, {\sl `Spacelike Singularities and String Theory'}, preprint
EFI-94-62,
hep-th/9412074.}.

It is somewhat ironic that although ultimately we hope to understand the
quantum gravity aspects of curved spacetime through string theory,
nearly all of our progress in this area so far has been in understanding
classical string theory. At first sight,
this might appear to be a disappointment, but that is not the
case. The simple fact that we have replaced point particle theory by a
theory of an extended object, \ie the  string, has important consequences.
Immediately we allow the stringy nature of our fundamental theory to become
relevant  we receive
corrections to the field equations of our particle theory. Expressed as a
perturbative series in $\alpha^\prime$ (the inverse string tension) this
infinite series of corrections---the $\beta$--function equations---invite the
possibility that even this classical theory might tell us a great deal about
the nature of spacetime singularities, etcetera. This is because it is in
precisely
in regions of high curvature that these classical stringy corrections to our
classical particle theory understanding of spacetime  are non--negligible.

The program of finding solutions to the leading order $\beta$--function
equations has considerable momentum\foot{See the review
of ref.\ref\dark{G. T. Horowitz, in the  proceedings of the
 Trieste Spring School, {\sl ``String theory and Quantum Gravity
'92''}, World Scientific 1993, hep-th/9210119.}\ for a
summary of some of the progress in this area.}.
Only limited efforts have been made in studying the effects of
next--to--leading order corrections in the $\beta$--function
equations\ref\firstorder{R.C. Myers, {\sl Nucl. Phys.} {\bf B289} (1987)
701\semi C.G. Callan, R.C. Myers and M.J. Perry, {\sl Nucl. Phys.}
{\bf B311} (1989) 673\semi
B.A. Campbell, N. Kaloper and K.A. Olive, {\sl Phys. Lett.}
{\bf B285} (1992) 199\semi
M.J. Duncan, N. Kaloper and K.A. Olive, {\sl Nucl. Phys.}
{\bf B387} (1992) 215\semi
M. Natsuume, {\sl Phys. Rev.} {\bf D50} (1994) 3949, hep-th/9406079.}.
Instead much progress has come in the investigation of cases
where this brute force approach can be side--stepped.
These include exact solutions, for which all of
$\alpha^\prime$ `corrections' vanish\ref\exact{R.C. Myers,
{\sl Phys. Lett.} {\bf B199} (1987)
371\semi
I. Antoniadis, C. Bachas, J. Ellis and D.V. Nanopoulos,
{\sl Phys. Lett.} {\bf B211} (1988) 393; {\sl Nucl. Phys.} {\bf B328}
(1989) 117\semi
R. G\"uven, {\sl Phys. Lett.} {\bf B191} (1987) 275\semi
D. Amati and C. Klim\v c\'\i k, {\sl Phys. Lett.} {\bf B219} (1989) 443\semi
G. Horowitz and A. Steif,  {\sl Phys. Rev. Lett.} {\bf 64} (1990) 260\semi
G. Horowitz and A. Steif, {\sl Phys. Rev.} {\bf D42} (1990) 1950\semi
 G.T. Horowitz, in: {\it
 Strings '90}, (eds. R. Arnowitt et. al.)
 World Scientific (1991)\semi
H. de Vega and N. Sanchez, {\sl Phys. Lett.} {\bf B244} (1990) 215\semi
G.T. Horowitz and A. A. Tseytlin, {\sl Phys. Rev.} {\bf D50} (1994) 5204,
hep-th/9406067\semi
G.T. Horowitz and A.A. Tseytlin, {\sl `A new class of exact solutions in
string theory'}, preprint IMPERIAL-TP-93-94-54 (1994),
hep-th/9409021.}\ref\except{G. T. Horowitz and A. A. Tseytlin,
{\sl `Extremal black holes as exact string solutions'}, preprint
IMPERIAL-TP-93-94-51 (1994), hep-th/9408040.\semi
R. Kallosh and %r extra and ,
T. Ortin,  {\sl Phys. Rev.}  {\bf D50} (1994) 7123, hep-th/9409060.}, and
conformal
field theory (CFT) methods.
The latter, which shall be considered in this paper, may be regarded as
solutions of string theory not only to all orders in $\alpha^\prime$,
but also incorporating effects non--perturbative in $\alpha^\prime$.

The study of black hole physics  with conformal field theories first arose
in the  the pioneering work
that was presented in ref.\ref\witten{E. Witten, {\sl Phys. Rev.} {\bf D44}
(1991) 314.}, showing that the $SL(2,\rline)/U(1)$  coset
  is
the  classical solution of string theory in a bosonic
two--dimensional black hole
background\foot{Prior to this, work regarding cosets based on
non--compact groups as candidates for curved spacetime string
theory backgrounds was presented in
ref.\ref\bars{I. Bars, {\sl Nucl. Phys. } {\bf B334}
(1990) 125\semi
I. Bars and D. Nemeschansky, {\sl Nucl. Phys. } {\bf B348} (1991) 89.}.}.
This result stimulated the discovery of many %r word
new CFT's which correspond to interesting gravitational backgrounds
in diverse dimensions. Our attention shall be focused
upon four--dimensional backgrounds for self--explanatory reasons.
Unfortunately, no exact CFT solution providing
a complete description of a four-dimensional black hole (including
the asymptotically flat regions) has yet been
constructed\foot{See, however, refs.\except\ for complete black hole solutions
to string theory which receive no $\alpha^\prime$ corrections. These are exact
at the level of the sigma--model description of the classical string theory.
Having such solutions is a very significant advance, although
the question of how to
in general construct their description as a   CFT (\ie\ construct the spectrum
of vertex operators and their correlations) still remains.
Such a situation is familiar in the case of some instanton solutions of
heterotic string theory\ref\chs{C. G. Callan, Jr., J. A. Harvey, A. Strominger,
in the proceedings of the Trieste Spring School,
 {\sl ``String Theory and Quantum Gravity '91''},
World Scientific 1992, hep-th/9112030\semi
C. G. Callan, Jr., J. A. Harvey, A. Strominger,
{\sl Nucl. Phys. } {\bf B359} (1991) 611.},
where the sigma models description is known to be
exact in $\alpha^\prime$, but only the throat  limit of a special
 case has been given a CFT description. However, see
ref.\ref\edADHM{E. Witten, {\sl `Sigma Models And The ADHM Construction Of
Instantons'}, Institute For Advanced Study Preprint
IASSNS-HEP-94-75, hep-th/9410052.}\
for a different approach to the problem.}. However
amongst the many four-dimensional solutions constructed in this
way\ref\fourd{I. Bars and K. Sfetsos, {\sl Phys. Lett.} {\bf B277}
(1992) 269, hep-th/9111040\semi C.R. Nappi and E. Witten, {\sl Phys. Lett.}
{\bf B293} (1992) 309, hep-th/9206078; {\sl Phys. Rev. Lett.} {\bf 71}
(1993) 3751, hep-th/9310112\semi P. Horava, {\sl Phys. Lett.}
{\bf B278} (1992) 101, hep-th/9110067\semi D. Gershon,
{\sl Nucl. Phys.} {\bf B421}
(1994) 80, hep-th/9311122; {\sl Phys. Rev.} {\bf D49} (1994) 999,
hep-th/9210160; preprint TAUP-1937-91, hep-th/9202005.},
a number of solutions corresponding to the
`horizon $+$ throat' region\foot{See section 4.2 for a description of these
regions.}\
of extremal black holes exist\ref\gps{S. B. Giddings,
J. Polchinski and A. Strominger,
{\sl Phys. Rev.} {\bf D48} (1993) 5748, hep-th/9305083.}\ref\lowe{D. A. Lowe
and A. Strominger {\sl Phys. Rev. Lett.}{\bf 73} (1994) 1468,
hep-th/9403186.}\ref\cvj{C. V. Johnson,  {\sl Phys. Rev.} {\bf D50} (1994)
4032, hep-th/9403192.}\ref\ghpss{S. B. Giddings, J. A. Harvey, J. G.
Polchinski, S. H. Shenker and A. Strominger,
{\sl Phys. Rev.}  {\bf D50} (1994) 6422, hep-th/9309152.}.

\def\SM{sigma--model}
In ref.\gps, a CFT was presented as
a solution of heterotic
string theory. The low energy limit of this CFT is a \SM\
whose
couplings correspond to the `horizon $+$ throat' region of the
 extremal limit of the magnetically
charged black hole solution of ref.\ref\charged{G.W.
Gibbons, {\sl Nucl. Phys.} {\bf B207} (1982) 337\semi
G.W. Gibbons and K. Maeda, {\sl Nucl. Phys.} {\bf B298} (1988) 741.}
\ref\chargedtwo{D. Garfinkle, G. Horowitz and A. Strominger, {\sl Phys. Rev.}
{\bf D43} (1991) 3140; {\bf D45} (1992) 3888(E).}. %r split ref
The CFT was described as the
product of the $SL(2,\rline)/ U(1)$ coset (supersymmetrised) with an asymmetric
orbifold of affine $SU(2)$. Indeed, the  whole spacetime solution inherits this
product form, the angular and time--radius sectors being completely decoupled.

Refs.\lowe\ and \cvj\ provide two distinct generalizations
of this solution. First,
ref.\lowe\ performed an analogous orbifolding of affine
$SL(2,\rline)$ to construct a family of
four-dimensional black hole solutions with both electric and
magnetic charges.
In ref.\cvj, the solution of ref.\gps\ was interpreted as an
example of a class of  conformal field theories which are called
`heterotic coset models'\ref\hetcosone{C. V. Johnson, {\sl `Heterotic Coset
Models'}, PUPT--1499, to appear in {\sl Mod. Phys. Lett.} {\bf A},
hep-th/9409062.}. These CFT's combine the
ingredients of a WZW model, left-- and right--moving fermions, and
non--dynamical world--sheet gauge fields so as to provide
background solutions of
heterotic string theory. These heterotic coset
constructions furnish a powerful means of generalising the
 solution of ref.\gps.
Indeed, it is straightforward to move beyond the direct product
form of the original construction and, for example, to produce non--trivial
mixing of the time--radius %r comma busters ,
and angular sectors.
Ref.\cvj\ presented one such example of the latter, and it was
conjectured that the new background was a stringy cousin of the
Taub--NUT solution\ref\TaubNUT{A.H. Taub, {\sl Ann. Math.} {\bf 53}
(1951) 472\semi E. Newman, L. Tamborino and T. Unti, {\sl J. Math. Phys.}
{\bf 4} (1963) 915.} of Einstein's equations, possessing non--trivial
dilaton and axion fields and with both electric and magnetic charges.

The latter conjecture
was confirmed in ref.\ref\us{C. V. Johnson and R. C. Myers,  {\sl Phys.
Rev.} {\bf D50} (1994) 6512, hep-th/9409069.}. There, stringy
solution generating techniques, namely
$O(d,d+p)$\ref\odd{S. Cecotti, S.
Ferrara and L. Girardello, {\sl Nucl. Phys.} {\bf B308}
(1988) 436\semi
K. Meissner and G. Veneziano, {\sl Phys. Lett.} {\bf B267} (1991) 33\semi
A. Sen, {\sl Phys. Lett.} {\bf B274} (1992) 34;
{\sl Phys. Lett.} {\bf B271} (1991) 295\semi
M. Gasperini, J. Maharana and G. Veneziano, {\sl Phys. Lett.} {\bf B272}
(1991) 277\semi
S. Hassan and A. Sen, {\sl Nucl. Phys.} {\bf B375} (1992) 103.}\  and
$SL(2,\rline)$\ref\sltr{A. Shapere,
S. Trivedi and F. Wilczek, {\sl Mod. Phys. Lett.} {\bf A6}
(1991) 2677\semi
A. Sen, {\sl Nucl. Phys.} {\bf B404} (1993) 109.}\ transformations,
were applied to the Taub--NUT solution of General Relativity
to construct a
leading order Taub--NUT dyon solution of low--energy heterotic
string theory. In the extremal limit, the fields in the `horizon $+$ throat'
 region
of the dyon were shown to match precisely
the background fields of the heterotic coset constructed in
ref.\cvj. At the same time, this Taub-NUT dyon
was also displayed in refs.\ref\them{R. Kallosh, D. Kastor, T.
Ortin and T. Torma, {\sl Phys. Rev.} {\bf
D50} (1994) 6374, hep-th/9406059.}\ and \ref\galtsov{D. V. Gal'tsov and
O. V. Kechkin, {\sl `Ehlers--Harrison--Type Transformations in  Dilaton--Axion
Gravity,'}, preprint MSU-DTP-94-2-REV, July 1994,
 hep-th/9407155.}\ as a special case of larger
families of leading order solutions constructed there.
In particular, ref.\galtsov\ constructed a family of solutions
which, as well as a NUT parameter, included an angular momentum
parameter. The latter represents a new non--trivial mixing of the
coordinates, and so the question naturally arises as to how
one can construct a conformal field
theory which describes this stringy Kerr--Taub--NUT solution to all orders in
the $\alpha^\prime$ expansion.

This is the question which we address in the present paper.
In the next section, we present a candidate for the CFT,
which is constructed using heterotic
coset techniques\cvj\hetcosone.
Section~3 extracts the low--energy content of this CFT,
exhibiting the spacetime metric, gauge fields, axion and dilaton, and shows
that the metric contains  a rotation parameter.
Section~4 examines the leading order Kerr--Taub--NUT solution of
heterotic string theory\galtsov\
and shows that in an  extremal limit its `horizon $+$ throat' sector
coincides with the solution in section~3.
%r some descriptive words which we had talked about
This limit has the interesting property that while the throat
region possesses a characteristic angular velocity, the angular
momentum vanishes in the asymptotically flat region.
Section~5 briefly presents a  generalisation of the CFT of section~2.
%r another descriptive sentence
The introduction of a new gauging parameter in this CFT produces
interesting modifications of the spacetime geometry, but we lack
a complete physical interpretation of this new parameter.
Section~6 summarises and concludes the paper.

\def\ap{\alpha^\prime}

\section{A Conformal Field Theory}

\subsection{Heterotic Sigma Models}
We begin with a two--dimensional \SM\ which describes the
propagation of heterotic strings in a non--trivial
background field configuration\ref\curt{C.G. Callan, E.J. Martinec,
M.J. Perry and D. Friedan, {\sl Nucl. Phys.} {\bf B262} (1985) 593.}:
\eqn\heterotic{\eqalign{I={1\over4\pi\alpha^\prime}\int d^2z&
\left[\left\{G_{\mu\nu}(X)+B_{\mu\nu}(X)\right\}\d X^\mu\db X^\nu
+{\alpha^\prime\over4}\Phi(X)R^{(2)}
\right]\cr
+{i\over\pi\alpha^\prime}\int d^2z&
\biggl[\lambda^a_R(\db-i\Omega_{\mu ab}(X)\db X^\mu)\lambda^b_R
+\lambda^\alpha_L(\d-{i}A_{\mu\alpha\beta}(X)\d X^\mu)
\lambda^\beta_L\cr
&\hskip5.0cm+{{2}}F_{\mu\nu\alpha\beta}(X)\Psi^\mu_R\Psi^\nu_R
\lambda^\alpha_L\lambda^\beta_L\biggr] .}}
Here, $(a,b)$ indicate tangent space indices
on the background field spin connection $\Omega$, and $(\alpha,\beta)$
are  current algebra indices on the spacetime gauge field $A$.
In a consistent string theory background,
the metric, dilaton, antisymmetric tensor and gauge fields are
balanced against each other in such a way so as to ensure
that the \SM\ is Weyl invariant. Thus demanding that the
\SM\ $\beta$--functions vanish yields the equations of motion for the
background fields\curt. Given the two-dimensional quantum field
theory defined by the action \heterotic, the $\beta$--functions
may be calculated perturbatively in the quantum loop expansion
in which $\alpha^\prime$ plays the role of $\hbar$.\foot{We will
set $\ap=2$ for the remainder of our discussion, except where
explicitly indicated.}

To go beyond this perturbative expansion all the way to defining a conformal
field theory, we note there are only a few ways known to define conformal field
theories by explicit Lagrangian methods. In addition to free massless field
theories, we have \wzw\
models\ref\wzws{E. Witten, {\sl Comm. Math. Phys.} {\bf 92} (1984) 455\semi
S. P. Novikov, {\sl Ups. Mat. Nauk.} {\bf 37}, (1982) 3\semi
V. G. Knizhnik and  A. B. Zamolodchikov, {\sl Nucl. Phys. } {\bf B 247} (1984)
83\semi
D. Gepner and E. Witten, {\sl Nucl. Phys.} {\bf B278} (1986) 493.}\
and their variants, and the list is largely
complete.
The conformal field theory which we will construct here
is a heterotic coset model\cvj\hetcosone\ which combines both of
these types of CFT. We introduce
a WZW model, some right--moving fermions to produce world sheet supersymmetry,
and some left--moving fermions. The background fields
$G_{\mu\nu}(X),$ $B_{\mu\nu}(X),$ $A_\mu(X)$ and $\Phi(X)$ are determined
by how we choose to couple these three ingredients on the world sheet.
We are guided by two `principles' in our construction.
The first is to  preserve as many of the spacetime symmetries as possible
in accord with the symmetries of the leading order low energy solution.
The second is, of course, to ensure conformal invariance in the final model.

\subsection{Metrics and the WZW Sector.}
In the dyonic Taub--NUT example of ref.\cvj\ (which contains the magnetic\gps\
and dyonic\lowe\ black holes as  special cases), the construction
begins with a WZW model based upon
$G=SL(2,\rline)\times SU(2)$. Alone, this would describe strings
propagating on a six--dimensional product manifold given by the group $G$ with
non--trivial metric and
antisymmetric tensor fields. It also possesses
a large affine $G_L\times G_R$ symmetry, acting as
\def\zb{{\bar{z}}}
\eqn\large{\eqalign{
g_1\to& g_1^L(z)g_1^{\phantom R}g_1^R(\zb)\cr
g_2\to& g_2^L(z)g_2^{\phantom R}g_2^R(\zb)\cr
{\rm for}\qquad g_1^{\phantom R},g_1^L,g_1^R\in& SL(2,\rline),
\qquad\qquad g_2^{\phantom R},g_2^L,g_2^R\in SU(2).
}}
Naively the idea is to restrict string propagation on the whole of this
manifold to a submanifold by gauging away some of the
two-dimensional \SM's symmetry \large.
With care, we can choose our gaugings such that we preserve some of the
desirable spacetime symmetries.

The following discussion will be facilitated by giving an explicit
parameterisation for the group elements. The $SU(2)$ manifold is $S^3$,
and we may choose a parameterisation in terms of Euler coordinates
\eqn\euler{g_2=\e{i\phi\sigma_3/2}\e{i\theta\sigma_2/2}\e{i\psi\sigma_3/2}
=\pmatrix{
\e{{i\over2}\phi_+}\cos{\theta\over2}&
\e{{i\over2}\phi_-}\sin{\theta\over2}\cr
-\e{-{i\over2}\phi_-}\sin{\theta\over2}&
\e{-{i\over2}\phi_+}\cos{\theta\over2}\cr
},}
where the $\sigma_i$ are the Pauli matrices and
\eqn\ranges{\phi_\pm\equiv\phi\pm\psi,\jump
0\leq\theta\leq\pi,\jump0\leq\phi\leq2\pi,\jump0\leq\psi\leq4\pi.}
For $SL(2,\rline)$, we choose
\eqn\euleri{g_1=\e{\tl\sigma_3/2}\e   {\sigma\sigma_1/2}\e{\tr\sigma_3/2}
=\pmatrix{
\e{{t_+\over2} }\cosh   {\sigma\over2}&
\e{{t_-\over2} }\sinh   {\sigma\over2}\cr
\e{-{t_-\over2} }\sinh   {\sigma\over2}&
\e{-{t_+\over2} }\cosh   {\sigma\over2}\cr
},}
with
\eqn\rangesi{t_\pm\equiv\tl\pm\tr,\jump
0\leq\sigma\leq\infty,\jump-\infty\leq\tl\leq\infty,
\jump-\infty\leq\tr\leq\infty.}
In the final model, the time and
radial spacetime coordinates come from the non--compact $SL(2,\rline)$
and the angular coordinates from the $SU(2)$.

\def\co{coordinate}
\subsubsection{Rotational Symmetry}
One of the key symmetries of the spacetime metrics of refs.\gps\lowe\cvj\ is
rotational invariance. Leaving the $SU(2)_L$ of the
$G_L\times G_R$ symmetry group intact results
in the rotational symmetry of the final spacetime background.
The approach is as follows: The $SU(2)$
manifold, $S^3$ is a   $U(1)$ fibre
bundle of $S^1$ over $S^2$, the Hopf fibration.
By gauging the $U(1)$ tranformations
\eqn\rightu{U(1)_R:\jump g_2\to g_2\e{i\epsilon\sigma_3/2}}
which act by right multiplication with $\epsilon(z,\zb)$
(\ie translations in $\psi$), only the $S^2$ remains with
\co s $(\theta,\phi)$. The $SU(2)_L$ symmetry, \ie $g_2^{\phantom L}\to
g^L_2(z)g_2$,
is preserved by  this gauging, and  acts as spacetime
rotations on the remaining spatial coordinates.

Part of the motivation of ref.\cvj\ was to mix the time--radius  and angular
sectors to obtain a non--product, but still rotationally invariant
background. The final symmetries which were gauged were:
\eqn\gaugingsone{U(1)_A\times U(1)_B:\left\{\jump\eqalign{
g_1\to&\e{\epsilon_A\sigma_3/2}g_1
\e{(\delta\epsilon_A+\lambda\epsilon_B)\sigma_3/2}\cr
g_2\to&g_2\e{i\epsilon_B\sigma_3/2}   }\right.}
with $\epsilon_A(z,\zb)$ and $\epsilon_B(z,\zb)$.
At this point in the  discussion, $\delta$ and
$\lambda$ are arbitrary constants.
A coupling between the $SL(2,\rline)$ and $SU(2)$ sectors is
achieved with non--zero $\lambda$. The
 $U(1)_B$ acts on  the $\psi$ and
$\tr$ fields. Since the latter is related to what becomes
the time \co\ in the final solution, the rotations induced
by $SU(2)_L$ will act on time as well as the angular coordinates.
Thus one loses spherical symmetry of the spacetime, in the conventional
sense, as can be seen from the final stringy metric\cvj:
\eqn\nutty{dS^2\sim d\sigma^2-f(\sigma)[dt+2\lambda A_\phi^M(\theta)d\phi]^2+
d\theta^2+\sin^2\theta d\phi^2}
where $2A_\phi^M(\theta)=
\pm1-\cos\theta$. The $\pm$ choice refers to either
%r old: the North or South pole of the $S^2$.
%r new:
the Northern or Southern hemispheres of the $S^2$ (\ie $\theta\le{\pi\over2}$
and $\theta\ge{\pi\over2}$, respectively).
Here the gauging parameter
$\lambda$ has become the NUT parameter in the final Taub--NUT
metric\TaubNUT. It is well known though, that the Taub--NUT space
is $SO(3)$ rotation invariant, but that these symmetry transformations
act on $t$ as well as the angular coordinates\ref\misner{C.W.
Misner, {\sl J. Math. Phys.} {\bf 4} (1963) 92\semi
C.W. Misner and A.H. Taub, {\sl Sov. Phys. JETP} {\bf 28} (1969) 122\semi
C.W. Misner in {\sl Relativity Theory and Astrophysics I: Relativity
and Cosmology}, ed. J. Ehlers, Lectures in Applied Mathematics {\bf 8}
(American Mathematical Society, 1967) 160.}\
in order to preserve the form of the differential
$dt+2\lambda A_\phi^M(\theta)d\phi$.

Another interesting feature of the Taub--NUT space is that the
surfaces of constant radius have the topology of three--spheres
in which the time direction has periodicity $4\pi\lambda$.
Thus $t$ becomes the $S^1$ fibre in the Hopf fibration over the $S^2$
with \co s $(\theta,\phi)$. The \co\ $t$ is given a period
($4\pi\lambda$)  by studying the action of rotations on
the metric \nutty\ \misner.

We could have deduced this periodicity of $t$ in advance by examination of the
gauging \gaugingsone.
 Consider the $U(1)_B$ transformation with $\epsilon_B=4\pi$.
This acts with the identity on the $SU(2)$ space (alternatively,
$\psi$ is shifted by a full period, \ie $\psi\to\psi+4\pi$),
while in $SL(2,\rline)$ it translates $t_R\to\tr+4\pi\lambda$.
Hence gauging $U(1)_B$ identifies $t_R\simeq\tr+4\pi\lambda$, and
the same periodicity is imposed on the final time coordinate
(\eg consider gauge fixing  $\tl=0=\psi$, which leaves $t=\tr$
--- see below).
The $S^3$ topology of constant $\sigma$ surfaces is most readily
evident with the gauge fixing $\tl=0=\tr$  so that
 the ($t,\theta,\phi$) surfaces inherit the $S^3$ topology of
the underlying $SU(2)$ space, with $t=\lambda\psi$.

For the rest of the gauging \gaugingsone, $U(1)_A$ is a non--diagonal
generalisation of the gauging
used in ref.\witten\ for the two-dimensional black hole.
This construction on its own (\ie with $\lambda=0$, and neglecting the $SU(2)$
sector) was shown in
ref.\cvj\ to produce charged two-dimensional black
hole solutions of heterotic string theory. In the present construction,
it will contribute to the electric charge of the final dyon.

\subsubsection{Rotation}
In the metric for a dyon which rotates about the $\phi$--axis, there
must be a new coupling between the $t$ and $\phi$ \co s beyond
that appearing in \nutty. Such a coupling will be
parameterised by the
angular velocity, and will be further earmarked  by
the fact that it breaks the rotational symmetry. The latter indicates that the
$SU(2)_L$ should not be preserved in our construction, and this singles
out a unique (up to
scalings) modification of the gaugings \gaugingsone\ as a candidate:
\eqn\gaugingstwo{U(1)_A\times U(1)_B:\left\{\jump\eqalign{
g_1\to&\e{\epsilon_A\sigma_3/2}g_1
\e{(\delta\epsilon_A+\lambda\epsilon_B)\sigma_3/2}\cr
g_2\to&\e{i\tau\epsilon_A\sigma_3/2}g_2\e{i\epsilon_B\sigma_3/2}.}\right.}
With non--zero $\tau$, %r new comma
$U(1)_A$ will introduce a new $t$--$\phi$ coupling which
breaks the rotational symmetry. The parameter
$\tau$ should then be related to the angular velocity.

To this point, all of our considerations have
concentrated upon the possible geometry
which we might extract as submanifolds of $G=SL(2,\rline)\times SU(2)$ without
much concern for whether such a conformal field theory can exist.
Indeed, if the WZW model for $G$ was the only contribution to the
world sheet action, there would be cause for dismay,
for upon  introducing world--sheet
gauge fields $(A^A_z,A^A_\zb)$ and
$(A^B_z,A^B_\zb)$ to enforce \gaugingstwo\ (or \gaugingsone) as a local
symmetry, we would find that our attempts to construct a gauge invariant model
are thwarted by the Wess--Zumino term of the WZW: We have chosen an `anomalous
subgroup' of the WZW model to gauge. Confident that we can fix this problem
later\cvj\hetcosone, let us parameterise our failure to find this gauge theory
thus far.  We choose to
write an extension $I(g_1,g_2,A^A,A^B)$ to the WZW model which, upon variation
of the fields according to \gaugingstwo, (and the gauge fields as $\delta
A=d\epsilon$) produces terms which do not depend upon $g_1$ or $g_2$. Such an
action is unique\ref\edholo{E. Witten, {\sl Comm. Math. Phys.} {\bf 144 } 189
(1992).}, and we shall postpone writing it until a little later.
However, we list the `classical anomaly' terms which the variation produces:
\def\anom#1#2#3{{#1\over4\pi}\int \!d^2z\,\epsilon^#2F^#3_{z\zb}}
\def\anoml#1#2#3{#1{1\over4\pi}\int \!d^2z\,\epsilon^#2F^#3_{z\zb}}
\eqn\wanomaly{\eqalign{
\anoml{(k_1(\delta^2-1)&-k_2\tau^2)}{A}{A}+\anom{k_1\delta\lambda}{A}{B}\cr
&+\anom{k_1\delta\lambda}{B}{A}+\anoml{(k_1\lambda^2+k_2)}{B}{B},
}}
where $F_{z\zb}\equiv \d\Azb-\db\Az,$ and $k_1$ and $k_2$ are the levels of
$SL(2,\rline)$ and $SU(2)$ respectively.
%r old: We have reversed the standard sign of $k_1$, \ie $k_1>0$,
%r to achieve a signature of $(-++)$ for the $SL(2,\rline)$ manifold\witten.
%r new:
We have reversed the standard sign conventions for $k_1$, \ie $k_1>0$
yields a $(-++)$ signature on the $SL(2,\rline)$ manifold\witten.
Now, let us move on to consider the other sectors of the theory
which also contribute to the full heterotic coset model.

%r\vfill\eject
\subsection{The Fermionic Sectors}
\subsubsection{Right--Movers}
First we need a family of right--moving fermions which are arranged to be
supersymmetric with the right--moving degrees of freedom of the spacetime
coordinates. Such a requirement is easy to satisfy. The Weyl fermionic field
$\Psi_R$ takes values in the orthogonal complement of Lie $H$ in Lie $G$,
where $H=U(1)_A\times U(1)_B$ and $G=SL(2,\rline)\times SU(2)$. Hence
we introduce four independent components $\psi_R^a$ where $a=1\ldots4$ are
tangent space indices on the coset manifold. By minimally coupling
them to the adjoint action of the gauge fields, world sheet supersymmetry is
ensured as will be discussed %r shown - it seems we don't explicitly show it
later when we exhibit the complete model.

Due to the chiral nature of the fermions, their minimal couplings to the gauge
fields, although classically gauge invariant, will produce chiral anomalies at
one loop. In an appropriate normalisation,  these are:
\eqn\Ranomaly{\eqalign{
&\anom{2\delta^2}{A}{A}+\anom{2\delta\lambda}{A}{B}\cr
&+\anom{2\delta\lambda}{B}{A}+\anoml{2(1+\lambda^2)}{B}{B}.
}}

\subsubsection{Left--Movers}
We introduce some left--moving fermions from the current algebra fermions
which carry the spacetime gauge group of the heterotic string.
They will couple to the rest of the model through their interactions
with the world-sheet gauge fields.
Without the constraints of attaining world sheet
supersymmetry, we may introduce these fermions with some freedom. Let us choose
four left movers $\lambda_L^\alpha$ arranged as a column vector $\Lambda_L$,
and minimally couple them to the gauge fields with generators:
\eqn\generators{{\hat Q}_A=\pmatrix{0&Q_A&0&0\cr -Q_A&0&0&0\cr0&0&0&P_A\cr
0&0&-P_A&0},{\hat Q}_B=
\pmatrix{0&Q_B&0&0\cr -Q_B&0&0&0\cr0&0&0&P_B\cr 0&0&-P_B&0},}
acting in the fundamental representation of $SO(4)$, as its
maximal torus subgroup,
\ie under infinitesimal gauge transformations \gaugingstwo,
$\delta\Lambda_L=i(\epsilon_A{\hat Q}_A+\epsilon_B{\hat Q}_B)\Lambda_L$.

Note that many other choices can be made at this point.
For generic values of the couplings $(\lambda,\delta,\tau,Q_A,Q_B,P_A,P_B)$,
the background space--time gauge fields fall in an Abelian
$U(1)\times U(1)$ subgroup, with identical
arrangements of the electric and magnetic charges in each factor.
One could have chosen to introduce a single pair of left--moving fermions
which would result in a single  background $U(1)$ gauge field.
This particular doubled arrangement was chosen here and in ref.\cvj\
because it allows a dyonic
model to be defined for arbitrary values of $\lambda$.
In particular when $\lambda$  vanishes,
both sets of couplings $(Q_A,Q_B,P_A,P_B)$ are required
to satisfy the anomaly cancelation conditions for the mixed ($AB=BA$) sector
while retaining charges other than the magnetic $Q_B$ --- see below.
For the purposes of comparison with the leading order spacetime
solution, though, we will only retain $(Q_A,Q_B)$ and hence a single  $U(1)$
background gauge group. The special case $\lambda=0$ will be mentioned
explicitly when necessary.

As before, the fermions will have chiral anomalies at one loop. These are:
\eqn\Lanomaly{\eqalign{
-&\anoml{2(Q_A^2+P_A^2)}{A}{A}-\anoml{2(Q_AQ_B+P_AP_B)}{A}{B}\cr
&-\anoml{2(Q_AQ_B+P_AP_B)}{B}{A}-\anoml{2(Q_B^2+P_B^2)}{B}{B}.
}}
As these are fermions of opposite chirality to the right movers, there is a
relative minus sign between \Ranomaly\ and \Lanomaly.

\def\hQ{{\hat Q}}
\def\hP{{\hat P}}

\subsection{A Consistent Model}
Combining all of the gauge anomaly terms, \wanomaly, \Ranomaly\
and \Lanomaly, we see that all of the anomalies cancel if
\eqn\cancel{\eqalign{k_1(\delta^2-1)-k_2\tau^2&=
2(Q_A^2+P_A^2-\delta^2)\cr
k_2+k_1\lambda^2&= 2(Q_B^2+P_B^2-(1+\lambda^2))\cr
k_1\delta\lambda&=2(Q_AQ_B+P_AP_B-\lambda\delta)\ .}}
Then the combination of the WZW model, and
the left-- and right--moving fermionic  sectors, all
coupled in the manner described above, will be gauge invariant and thus
describe a consistent conformal field theory.

The complete heterotic coset model is\cvj\hetcosone:
\eqn\totalaction{\eqalign{I=I_{WZW}
+&{k_1\over8\pi}
\int
d^2z\,\,\Biggl\{-2\left(\delta\Azb^A+\lambda\Azb^B\right)\Tr[\sigma_3g_1^{-1}\d
g_1]-2\Az^A\Tr[\sigma_3\db g_1g_1^{-1}]\cr
&\jump\jump+\Az^A\Azb^A\left(1+\delta^2+\delta
\Tr[\sigma_3g_1\sigma_3g_1^{-1}]\right)+
\lambda^2\Az^B\Azb^B\cr
&\jump\jump+\delta\lambda\Az^A\Azb^B+
\Az^B\Azb^A
\left(\delta\lambda+\lambda\Tr[\sigma_3g_1\sigma_3g_1^{-1}]\right)\Biggr\}\cr
+&{k_2\over8\pi}\int d^2z\,\, \Biggl\{2i\Azb^B\Tr[\sigma_3g_2^{-1}\d
g_2]+2i\tau\Az^A\Tr[\sigma_3\db g_2g_2^{-1}]\cr
&\jump\jump+\tau\Az^A\Azb^B\Tr[\sigma_3g_2\sigma_3g_2^{-1}]+\tau^2\Az^A\Azb^A
+\Az^B\Azb^B\Biggr\}\cr
-&{ik_1\over4\pi}\int d^2z\,\,
\Tr\left[\Psi_{R,1}(\db\Psi_{R,1}+(\delta\Azb^A
+\lambda\Azb^B)[\sigma_3/2,\Psi_{R,1}])\right]\cr
+&{ik_2\over4\pi}\int
d^2z\,\,\Tr\left[\Psi_{R,2}(\db\Psi_{R,2}+\Azb^B[i\sigma_3/2,\Psi_{R,2}])
\right]\cr
-&{ik_1\over4\pi}\int d^2z\,\,
\left(\Lambda^T_{L,1}[\d+i(Q_{A}\Az^A+Q_{B}\Az^B)\sigma_2]\Lambda_{L,1}
\right)\cr
+&{ik_2\over4\pi}\int
d^2z\,\,\left(\Lambda^T_{L,2}[\d+i(P_{A}
\Az^A+P_{B}\Az^B)\sigma_2]\Lambda_{L,2}
\right).
}}
Here, the fermions are decomposed as
\eqn\decompose{\eqalign{
\Psi_{R,1}&=\pmatrix{0&\psi^1_R\cr\psi^2_R&0}\cr
\Lambda_{L,1}&=\pmatrix{\lambda_L^1\cr\lambda_L^2}\cr} \qquad\qquad
\eqalign{
\Psi_{R,2}&=\pmatrix{0&\psi^3_R\cr\psi^4_R&0}\cr
\Lambda_{L,2}&=\pmatrix{\lambda_L^3\cr\lambda_L^4}\cr}}
where `1' and `2'
denote fermions coupling to the $SL(2,\rline)$ and $SU(2)$
sectors of the WZW model, respectively.

The model has invariance under the naive $(0,1)$ world--sheet
supersymmetry\ref\ed{E. Witten, {\sl Nucl. Phys.} {\bf B371} (1992) 191.}
\eqn\susan{\eqalign{\delta g_1&=i\epsilon g_1\Psi_{R,2}\qquad\qquad
\delta g_2=i\epsilon g_2\Psi_{R,2}\cr
\delta\Psi_{R,1}&=\epsilon\Pi_1\left(g_1^{-1}\d g_1+{1\over2}\Az^A
g_1^{-1}\sigma_3g_1+i\Psi_{R,1}\Psi_{R,1}
\right)\cr
\delta\Psi_{R,2}&=
\epsilon\Pi_2\left(g_2^{-1}\d g_2+i{\tau\over2}\Az^Ag_2^{-1}\sigma_3 g_2+
i\Psi_{R,2}\Psi_{R,2}\right)\cr
\delta A_i^A&=0=\delta A_i^B=\delta\Lambda_L,}}
(modulo equations of motion) which may be
verified by direct calculation. Here $\Pi_{1,2}$ projects back onto
 the orthogonal complement of Lie$H_{1,2}$ in Lie$G_{1,2}$.
One may also show  that this is
enhanced to $(0,2)$ supersymmetry since $G/H$ is a K\"ahler
coset\ed\ref\kazamasuzuki{Y. Kazama and H. Suzuki, {\sl Phys. Lett.} {\bf B216}
(1989) 122\semi Y. Kazama and H. Suzuki, {\sl Nucl. Phys.} {\bf B321} (1989)
232.}.

%r old:
%The final requirement is on the central charge of the theory. We shall adopt a
%central charge $c=6$ for our theory\foot{The $-2$ from gauging
%is canceled by the $+2$ from the four fermions.}\ assuming
%that there is an unspecified
%internal sector which produces a total of $c=15$ on the
%right and $c=26$ on the left.
%Thus we have
%\eqn\central{c={3k_1\over k_1-2}+{3k_2\over k_2+2}=6.}
%This results in the relation $k_1=k_2+4$ for the levels of each sector.
%r new:
The final requirement is on the central charge of the theory.
The central charge of our heterotic coset is\foot{The $-2$ from gauging
is canceled by the $+2$ from the four fermions.}
\eqn\central{c={3k_1\over k_1-2}+{3k_2\over k_2+2}\ .}
We assume that there is an unspecified
internal sector which produces a total of $c=15$ for the right--moving
sector and $c=26$ on the left. In order to make a comparison
with the low--energy solution of ref.\galtsov\ we will take the levels
$k_1,k_2\rightarrow\infty$ in which case $c\rightarrow6$. Note
that $c=6$ corresponds to the central charge of a weak field
four--dimensional heterotic string background as would
be required to describe the asymptotic regions of a black hole.
%r above sentence might need some work
We could also achieve $c=6$ for finite $k_1$ and $k_2$ by setting
$k_1=k_2+4$.

\section{The Low--Energy Limit}
The conformal field theory presented in the previous section represents
a solution of the classical heterotic string equations
to all orders in $\ap$ expansion, and including any non--perturbative
contributions as well.
To determine whether this conformal field theory makes contact with
the leading order Kerr--Taub--NUT solution of ref.\galtsov,
we shall extract from it the leading order background fields for our model.
Normally for gauged WZW models, the first step in this process
is to integrate out the world sheet gauge fields, which
appear only quadratically in the action.
Putting coordinates on the group manifold and gauge fixing
appropriately then yields the final background. We can apply
the same reasoning here, but we must first be careful. Recall
that we arrived at %r add "at"
a consistent model by canceling classical anomalies
of the bosonic WZW fields against one loop quantum anomalies of the fermions.
The first type appear explicitly in the action while
the second do not. Hence the coefficients
of the terms in the Lagrangian  quadratic
in the gauge fields do not account for the
fermion anomalies.

To surmount this problem we need to make these fermion contributions
appear at the classical level so that they explicitly enter the
world sheet action.
This is accomplished\cvj\ by bosonising the fermions.
Ref.\cvj\ constructed the bosonised theory for the fermions
of a similar model with
gauging \gaugingsone. Note that the extra parameter $\tau$ which
enters into the present gauging \gaugingstwo\ does not appear
amongst the fermion terms of our action \totalaction.
Therefore we can simply use the bosonic
theory of ref.\cvj:
\eqn\thebosons{\eqalign{I_B={1\over4\pi}\int d^2z\,\,\,
\Biggl\{&(\d\Phi_2-P_A\Az^A-(P_B+1)A^B_z)^2 \cr
+&(\d\Phi_1-(Q_B+\lambda)A^B_z-(Q_A+\delta)A^A_z)^2\cr
-&\Phi_1\Biggl[(Q_B-\lambda)F^B_{z\bar z}+(Q_A-\delta)F^A_{z\bar z}\Biggr]\cr
-&\Phi_2 \Biggl[(P_B-1)F^B_{z\bar z}+ P_AF^A_{z\bar z}\Biggr]\cr
+&\Biggl[\Azb^A\Az^B-\Az^A\Azb^B\Biggr]\Biggl[\delta Q_B-Q_A\lambda-P_A
\Biggr]\Biggr\}.}}
The  $U(1)_A\times U(1)_B$ action on the bosons $\Phi_1,\Phi_2$  is:
\eqn\acta{\delta\Phi_1=(Q_A+\delta)\epsilon_A+(Q_B+\lambda)
\epsilon_B\,\,\,\,\delta\Phi_2=P_A\epsilon_A+(P_B+1)\epsilon_B\ .}
With $\delta A^A_i=\partial_i\epsilon_A$ and
$\delta A^B_i=\partial_i\epsilon_B$ as usual, it is
simply verified that the action \thebosons\ yields the
fermion anomalies in eqs.\Ranomaly\  and \Lanomaly.

With this bosonised action replacing the fermionic terms in the
action \totalaction, the consistency of the gauging
is manifest at the classical level.
The gauge fields may be now integrated out by doing
a saddle point approximation for the corresponding Gaussian integrals.
The latter approximation is exact in the limit
$k_1\sim k_2\to\infty$, which is equivalent to an $\ap\to0$
%r new footnote
limit\foot{Note that one must also take the charges $Q_A,Q_B,P_A,P_B
\to\infty$ at the same time in order to preserve the anomaly cancelation
conditions \cancel.}.

Now using the parameterisation of the WZW model given in eqs.\euler\
and \euleri, the gauge transformations act by:
\eqn\transform{\eqalign{
&\sigma\to\sigma\qquad\theta\to\theta\cr
&\tl\to\tl+\epsilon_A\cr
&\tr\to\tr+\delta\epsilon_A+\lambda\epsilon_B\cr
&\psi\to\psi+\epsilon_B\cr
&\phi\to\phi+\tau\epsilon_A\cr
&\Phi_1\to\Phi_1+(Q_A+\delta)\epsilon_A+(Q_B+\lambda)\epsilon_B\cr
&\Phi_2\to\Phi_2+P_A\epsilon_A+(P_B+1)\epsilon_B.}}
One may verify that the action  is invariant under
these transformations (modulo the application of the
anomaly cancelation conditions \cancel).
Now we fix a gauge in which $\psi=t_L=0$, and denote $t_R=t$. (This differs
slightly from the gauge used in ref.\cvj. See subsection 5.2 for a discussion
of  the important relationship between world sheet gauge choices
and spacetime symmetries.)

Now that we have  performed the integration we  have arrived at a
bosonic action, but  we must restore the
fermions  before we interpret it as
a heterotic \SM\
and read off the background fields.  The
bosonised form has made the fermions' couplings appear at one order larger
 in
perturbation theory than they should be, (which was necessary before
integration to allow the Lagrangian to be sensitive to the fermion's
anomalies) and hence the
the background fields are presently shifted from
their correct values.
Now we must reintroduce the fermions into the resulting action
in order to correctly determine the background fields of the
heterotic sigma model.  This point is
discussed more in detail in ref.\cvj\ with
examples. Finally the dilaton coupling must be determined by an
evaluation of the fluctuation determinant for the integration
over the world sheet gauge fields\ref\dilatant{T.H. Buscher,
{\sl Phys. Lett.} {\bf B194} (1987) 59; {\sl Phys. Lett.} {\bf B201}
(1988) 466\semi E.B. Kiritsis, {\sl Mod. Phys. Lett.} {\bf A6}
(1991) 2871.}.

The final model is of the standard form \heterotic, and
the background fields  may be simply read off:
\eqn\thesolution{\eqalign{
dS^2&=k\Biggl[d\sigma^2+ d\theta^2
-\left({\sinh\sigma(dt-\lambda\cos\theta d\phi) \over
\cosh\sigma+\delta-\lambda\tau\cos\theta}\right)^2\cr
&\hskip1cm+\left({\sin\theta (\tau dt-(\cosh\sigma+\delta)d\phi)
\over\cosh\sigma+\delta-\lambda\tau\cos\theta}\right)^2\Biggr] \cr
\Phi-\Phi_0&=-\log[\cosh\sigma+\delta-\lambda\tau\cos\theta]\cr
B_{t\phi}&=-k{\tau+\lambda\cos\theta\cosh\sigma\over\cosh\sigma+
\delta-\lambda\tau\cos\theta}\cr
A_t&=-
{2\sqrt{2}(Q_A-\tau Q_B\cos\theta)\over\cosh\sigma+
     \delta-\lambda\tau\cos\theta}\cr
A_\phi&=-{2\sqrt{2}\cos\theta(Q_B(\cosh\sigma+\delta)-\lambda Q_A)
\over\cosh\sigma+\delta-\lambda\tau\cos\theta}}}
where $\Phi_0$ is a constant.
Note that we have not explicitly presented %r omitted
the second set of `mirror'
$U(1)$ gauge fields. These are identical
to gauge fields above with the replacement $Q\to P$.

\def\veps{\varepsilon}
{}From the antisymmetric tensor field and the gauge fields above, we can
calculate the scalar axion, $\rho$. This field is defined by the relation:
\eqn\defax{H_{\mu\nu\rho}=-\e{\Phi}\veps_{\mu\nu\rho\kappa}\nabla^\kappa
\rho}
where $\veps_{\mu\nu\rho\kappa}$ is the  volume form in four
dimensions\foot{Note that this definition is usually written in terms of the
Einstein metric, $g_{\mu\nu}=\e{-\Phi}G_{\mu\nu}$, but this metric
will play no role in the following.}. So here,
$\veps_{t\sigma\theta\phi}=\sqrt{-G}$, where $G$ is determinant
of the above sigma model metric. Also, the three--form $H$
is given by
\eqn\hfield{H_{\mu\nu\rho}=\partial_\mu B_{\nu\rho}+\partial_\nu B_{\rho\mu}+
\partial_\rho B_{\mu\nu}-\omega(A)_{\mu\nu\rho}}
where $\omega(A)$ is the Chern--Simons three--form for the gauge
sector\foot{We suppress the Chern--Simons
contribution from the Lorentz sector (\ie\ from the spin connection)
as higher order in the $\ap$ expansion --- see the next section.}:
\eqn\chern{\omega(A)_{\mu\nu\rho}={1\over4}\left(A_\mu F_{\nu\rho}+A_\nu
F_{\rho\mu}+A_\rho F_{\mu\nu} \right).}
\def\Ax{\rho} This gives:
\eqn\ouraxion{\Ax-\Ax_0=\e{-\Phi_0}(\lambda\cosh\sigma+\tau\cos\theta)}
with $\Ax_0$ is a  constant.

Recall that these fields are valid in the low energy limit,
\ie\ $k_1=k_2=k\to\infty$, which was required to
justify the saddle point approximation
for path integral over the world sheet gauge fields. In this limit,
the anomaly equations \cancel\ become:
\eqn\canceltwo{\eqalign{
{k\over2}={Q_A^2\over\delta^2-1-\tau^2}=
{Q_B^2\over1+\lambda^2}={Q_AQ_B\over\delta\lambda}\ ,
}}
where the extra spacetime $U(1)$'s have been deleted, \ie
$P_A=P_B=0$. Implicitly, the parameters appearing in our background fields
\thesolution\ obey these restrictions leaving three independent parameters.

As mentioned before the case $\lambda=0$ needs a little more
%r old footnote: care\foot{Note
%that $\delta^2>1+\tau^2$ if the charge $Q_A$ is to be real. Thus the possible
%complications of $\delta=0$ never arise here.}.
%r new footnote:
care\foot{Note that eq.\canceltwo\ requires $\delta^2>1+\tau^2$
if the charge $Q_A$ is to be real. Thus one cannot set $\delta=0$.
In fact, $\delta=0$ is not allowed within the full anomaly cancelation
conditions \cancel\ either.}.
%r i also rearranged the remainder of the paragraph as well. old:
%As can be seen from the full anomaly equations \cancel, in this case
%the mixed anomaly condition (\ie\ the $AB=BA$ sector)  is
%\eqn\abba{P_AP_B+Q_AQ_B=0.}
%In this case, in
%order to continue with a single spacetime $U(1)$ gauge group ($P_A=P_B=0$),
%$Q_B\neq0$ to satisfy the $BB$
%anomaly equation, and therefore the solution must
%be purely magnetically charged, \ie\ $Q_A=0$. Consequently for dyonic
%solutions at  $\lambda=0$ (with the arrangement of fermions presented in this
%paper), we need to retain the extra $U(1)$ sector (\ie\ $P_A,P_B\neq0$),
%with which we can satisfy \abba\ in a number of ways.
%r new arrangement:
In this case, solving the anomaly equations \canceltwo\ requires:
$\delta^2=1+\tau^2$, $Q_B^2={k\over2}$ and $Q_A=0$. Hence, the solution
would be purely magnetically charged.
If we consider the the full anomaly equations \cancel\
with vanishing $\lambda$ (and $P_A,P_B\neq0$),
the mixed anomaly condition (\ie\ the $AB=BA$ sector)  is
\eqn\abba{P_AP_B+Q_AQ_B=0.}
Consequently at $\lambda=0$, dyonic solutions are possible if
we retain the extra $U(1)$ sector (\ie\ $P_A,P_B\neq0$).
%r cut para. break
%r old stuff:
%So for the rest of the paper, the  dyonic $\lambda=0$ case is obtained by
%supplementing the fields in \thesolution\ (or some of the later
%solutions to be presented) with
%r new stuff:
So for the remainder of our discussion in the special case $\lambda=0$,
we will assume that the solution \thesolution\ is supplemented
with an extra set of component fields $A_t, A_\phi$
which are identical to those already listed except for the replacement of
$(Q_A,Q_B)$ by $(P_A,P_B)$. The low--energy anomaly equations are extended by
these charges in the obvious way, as can be seen from \cancel.

Notice that the axion field \ouraxion\ is unmodified by the extra $U(1)$, once
the extended low--energy anomaly equations are used.

\subsection{Some Spacetime Physics.}
Here, we study some of the physical properties of these leading
order spacetime fields. First,
notice that our solution is not asymptotically flat.
%r new sentence
Instead the size of the angular subspace becomes constant at
large $\sigma$. This fixed throat geometry %r extra words
is typical of the four--dimensional black hole solutions which have been
obtained as exact conformal field theories\gps\lowe\cvj.
It remains an open problem
to discover how  a (locally) asymptotically flat spacetime  may be
smoothly connected
onto the present  solution at the level of the conformal
field theory. See refs.\gps\ghpss\ for  discussions of such issues in the
context of
closely related models.

Our solution generalises the extremal dyonic
Taub--NUT solution  presented in ref.\cvj\ by the introduction of the
parameter $\tau$, and we would like to understand
its role more precisely. The background fields
\thesolution\ are invariant under time translations and axial
rotations. The corresponding Killing vectors are:
\eqn\killing{\xi^\mu\partial_\mu={\partial\,\over\partial t}
\hskip1cm{\rm and}
\hskip1cm \psi^\mu\partial_\mu={\partial\,\over\partial\phi}\ .}
Our solution has a Killing horizon, which is defined as
a surface upon which a (constant) linear combination of the
Killing vectors \killing\  is  null\ref\wald{R.M.~Wald, {\it General
Relativity} (University of Chicago Press, Chicago, 1984).}.
This surface corresponds
to $\sigma=0$, and the horizon generating Killing field is:
\eqn\nullvector{\chi^\mu\partial_\mu=
{\partial\,\over\partial t}
+\Omega_H{\partial\,\over\partial\phi} }
with
\eqn\angular{\Omega_H={\tau\over1+\delta}\ .}
It is easily verified that $\chi^\mu\chi_\mu|_{\sigma=0}
=\left|G_{tt}+2\Omega_HG_{t\phi}+\Omega_H^2G_{\phi\phi}\right|_{\sigma=0}=0$.
The standard interpretation of $\Omega_H$ is as the angular velocity
at the horizon\wald, and as anticipated it is
proportional to $\tau$.

The present coordinate system is not well--behaved at $\sigma=0$,
\eg the determinant of the metric vanishes there, and so
it is prudent to  make the following change of coordinates:
\def\tt{{\tilde{t}}}
\def\htt{{\hat{t}}}
\def\tphi{{\hat{\phi}}}
\eqn\prudent{\eqalign{u&=\sinh^2\sigma\cr
dt&=d\htt + {A(u)}du \cr
d\phi&=d\tphi + {B(u)}du\ ,}}
where
\eqn\ab{A(u)=-{\sqrt{(1+u)}+\delta\over2u\sqrt{(1+u)}}
\quad{\rm and}\quad
      B(x)=-{\tau\over2u\sqrt{(1+u)\ .}}}
The resulting metric is:
\eqn\formmetric{\eqalign{
dS^2=k&\Biggl[-{u\over U^2}(d\htt-\lambda\cos\theta d\tphi)^2 %r extra "+"
+{du\over U\sqrt{1+u}}(d\htt-\lambda\cos\theta d\tphi) \cr
&\qquad\hskip1cm+{\sin^2\theta\over U^2}
(\tau (d\htt-\lambda\cos\theta d\tphi)-
U\,d\tphi)^2+ d\theta^2\Biggr] \cr}}
where $U=\sqrt{1+u}+\delta-\lambda\tau\cos\theta$.
Hence
\eqn\detG{\sqrt{-G}=-{\sin\theta\over2\sqrt{1+u}
(\sqrt{1+u}+\delta-\lambda\tau\cos\theta)}}
and the coordinate singularity at $\sigma=0=u$ has been
eliminated.
In the new coordinates, the Killing
vectors are simply: $\xi^\mu\partial_\mu=\partial_\htt$ and
$\psi^\mu\partial_\mu=\partial_\tphi$. Since the new %r these
coordinates are perfectly regular at
the Killing horizon, they
extend the solution beyond the horizon to negative values of $u$.
In this region, a curvature singularity occurs at $u=-1$.
Further for $\lambda=0$, one can see that $u=0$ also plays the role of a future
event horizon\wald, in that
physical world--lines cannot escape from negative $u$ to positive $u$:
For any
{\it point} particle path $x^\mu(s)$, we demand that the local
four velocity is time--like, \ie $G_{\mu\nu}\dot x^\mu(s)
\dot x^\nu(s)\le0$. Now
for $\lambda=0$ and negative $u$, all of the contributions to the latter
expression are
positive definite except the $G_{\htt u}\dot\htt(s)\dot u(s)$ cross term.
Since $G_{\htt u}>0$ once behind the horizon at $u=0$,
any physical trajectory has ${\partial u\over\partial\htt}= %r extra stuff
\dot u(s)/\dot\htt(s)<0$, and moves towards
smaller values of $u$ and towards the singularity at $u=-1$.
Finally note that one can construct a coordinate patch which covers
the past event horizon in a non--singular way by changing the signs
of $A(u)$ and $B(u)$ in eq.\prudent.

With a non--vanishing value of $\tau$, the squared magnitude of
the time--translation Killing
vector, $\xi^\mu\xi_\mu=G_{tt}=G_{\htt\htt}$, reverses its sign
and becomes positive before one reaches %r old: outside of
the horizon at $u=0$.
Thus there exists in our solution an `ergosphere'
\eqn\ergo{0\leq u\leq\tau^2\sin^2\theta}
analogous to that of the Kerr solution of General Relativity.
Within this region because of the rotational
frame dragging, no particles can remain stationary even though
they are outside of the horizon\wald.

Another quantity of interest is $\kappa$, the surface gravity
of our solution, which may be defined by\wald:
\eqn\surfone{\left.\nabla_\nu(\chi^\mu\chi_\mu)\right|_H=
\left.-2\kappa\,\chi_\nu\right|_H\ .}
This quantity is related to the Hawking temperature of a black
hole\ref\radi{S.W.~Hawking, {\sl Comm. Math. Phys.} {\bf 43}
(1975) 199.}. %r took this out: via $T_H=\kappa/(2\pi)$
%r but added comments in discussion
Note that in the old coordinates, eq.\surfone\ is ill--defined,
but it easily evaluated in the new coordinate system \prudent\ yielding:
\eqn\surftwo{\kappa={1\over1+\delta}\ .}

As well as investigating the background geometry, we would like to
determine the electric and magnetic charges of our leading order
solution. Even though there is no asymptotically flat region, one
can expect to determine these charges through flux integrals
over the angular coordinates. For example, the magnetic charge
would be:
\eqn\magcharge{Q_M={1\over4\pi}\oint_{S^2} F}
where $F=dA$ is the electromagnetic field strength two--form.
We must remember, though, that for $\lambda\ne0$ the topology
of the solution changes so that $(\theta,\phi)$ do not define
a closed two-sphere. Hence this definition \magcharge\ may only
be applied for $\lambda=0$, in which case we find:
\eqn\magchargea{Q^{(1)}_M=2\sqrt{2}Q_B,}
from the fields in \thesolution\ and
\eqn\magchargeb{Q^{(2)}_M=2\sqrt{2}P_B,}
from the other $U(1)$ factor which we retain for dyonic solutions in the
$\lambda=0$ case.

\def\wF{\widetilde{F}}
\def\wX{\widetilde{X}}
A similar definition for the electric charge requires the
definition of a second closed two--form constructed from the field
strength tensor. In Einstein--Maxwell theory, the second form
is simply the Hodge dual of the field strength, $\wF$, and closure
is guaranteed by the equation of motion $d\wF=0$ or $\nabla^\nu
F_{\nu\mu}=0$. The leading
order heterotic string equations for the $U(1)$ gauge field may
be written
\eqn\gaugeom{\nabla^\nu X_{\nu\mu}=\nabla^\nu\left(e^{-\Phi}F_{\nu\mu}
+{1\over2}\rho\,\veps_{\nu\mu\alpha\beta}F^{\alpha\beta}\right)=0}
where as above $\veps_{\nu\mu\alpha\beta}$ is the volume four--form.
In terms of the dual of $X$, this equation of motion is $d\wX=0$
and so
\eqn\elecharge{Q_E={1\over4\pi}\oint_{S^2} \wX}
defines a topologically conserved charge. One also may verify that
for an asymptotically flat solution where $e^\Phi\to1+O(1/r)$
and $\rho\to O(1/r)$, $\wX\to\wF$ and the definition
\elecharge\ correctly yields the electric charge. For our present solution,
we find:
\eqn\elechargea{Q^{(1)}_E=2\sqrt{2}Q_A\ ,}
and from the other $U(1)$ factor,
\eqn\elechargeb{Q^{(2)}_E=2\sqrt{2}P_A\ .}
{}From \abba\ we see that we have at $\lambda=0$ the relation
$Q^{(1)}_MQ^{(1)}_E+Q^{(2)}_MQ^{(2)}_E=0,$ showing that we have now only three
independent charges in this special case, as could be anticipated by counting
 the number of parameters specified in the original model, and taking into
account the restrictions given by \cancel\ and \central.

\def\s{\sigma}
\def\th{\theta}
\def\ie{{\it i.e.,}\ }
\def\b{\beta}
\def\S{\Sigma}
\def\g{\gamma}
\def\w{\omega}
\def\W{\Omega}
\def\D{\Delta}
\def\a{\alpha}
\def\ll{\tilde\lambda}
\def\d{\delta}
\section{Kerr--Taub--NUT Dyons}
In the previous section, we firmly established that our solution
is rotating. It remains to be see whether
it precisely corresponds to the `horizon $+$ throat' region of the
Kerr--Taub--NUT dyon presented in ref.\galtsov.
First let us establish our conventions for the low energy
fields. We write the four--dimensional effective action
for the heterotic string as
\eqn\lowaction
{I=
\int d^4x\,\sqrt{-G}e^{-\Phi}\,\left(R(G)+(\nabla\Phi)^2-{1\over12}H^2
-{1\over8}F^2+\ldots\right)\ \ ,}
where the three--form $H$ is defined as in eq.\hfield\ (and
of course, $F_{\mu\nu}=\partial_\mu A_\nu-\partial_\nu A_\mu$).
The ellipsis indicates two sets of terms which may be ignored:
First, the full theory includes many other
massless fields ({\it e.g.,} more gauge fields, fermions, moduli fields,
etcetera), all of which may consistently be set to zero. Second, the $\ap$
expansion produces an infinite series of higher-derivative interactions,
whose contributions to the equations of motion
will be negligible for slowly varying fields.
In a standard normalization even the gauge kinetic terms
would appear amongst
the $O(\ap)$ interactions, but we have rescaled the gauge fields
by a factor of $1/\sqrt{\alpha'}$ in \lowaction. Therefore when
considering background solutions in the $\ap\to0$ limit,
we are thinking of them as carrying very
large (electric and magnetic) charges.
This explains why the gauge Chern--Simons contribution to the
three--form $H$ is included in eq.\hfield, but the Lorentz
Chern--Simons term is omitted.
Written in terms of the scalar axion \defax, this low energy action
becomes:
\eqn\lowacttwo{
\eqalign{I=\int d^4x\,\sqrt{-G}e^{-\Phi}\,&\left( R(G)+(\nabla\Phi)^2
-{1\over8}F^2\right.\cr
&\left.\ \ \ \ -{1\over2}e^{2\Phi}(\nabla\rho)^2
-{1\over16}e^{\Phi}\rho\,\epsilon^{\mu\nu\sigma\kappa}
F_{\mu\nu}F_{\sigma\kappa}+\ldots\right)\ .\cr}}

\subsection{The Low Energy Fields}
Ref.\galtsov\ constructs a Kerr--Taub--NUT dyon solution of
low energy heterotic string theory
as an example of the use of certain solution generating
techniques. This solution is a generalisation of
the stringy charged and rotating black hole presented in
ref.\ref\senkerr{A. Sen, {\sl Phys. Rev. Lett.}
{\bf 69} (1992) 1006, hep-th/9204046.}.
%r old stuff (part moved to footnote): Using the  convention that
%the \SM\  metric $G_{\mu\nu}=e^\Phi g_{\mu\nu}$, where $g_{\mu\nu}$ is the
%Einstein metric, the metric of ref.\galtsov\ becomes
%r new:
The \SM\ metric corresponding to the solution of ref.\galtsov\
is\foot{We use the convention that the \SM\ metric $G_{\mu\nu}=e^\Phi
g_{\mu\nu}$, where $g_{\mu\nu}$ is the Einstein metric.
We have also flipped the overall
sign of the metric presented in ref.\galtsov\ to produce
a $(-,+,+,+)$ signature. We also use different conventions
for a number of the other background fields: $\Phi=2\Phi^\prime$, %r extra 2
$\rho=-\kappa^\prime$ and $A_\mu=2\sqrt{2}A^\prime_\mu$ where the primed
fields are those used in ref.\galtsov.}:
\eqn\galtsovmetric{\eqalign{
dS^2=-{\W(\D-a^2\sin^2\theta)\over\S^2}&(d\tt-\w d\phi)^2\cr
+&\W\left({dr^2\over\D}+d\theta^2+{\D\sin^2\th\over\D-a^2\sin^2\th}d\phi^2
\right)}}
where
\def\td{\tilde{\delta}}
\eqn\now{\eqalign{
\D&=(r-r_-)(r-2M)+a^2-(N-N_-)^2\cr
\S&=r(r-r_-)+(a\cos\th+N)^2-N_-^2\cr
\w&={2\over a^2\sin^2\th-\D}\left\{N\D\cos\th+a\sin^2\th\left[
M(r-r_-)+N(N-N_-)\right]\right\}\cr
\W&=r^2-2Qy\,r+2(Nx-Py)\td+\td^2+Q^2x\cr
\td&=a\cos\th+N-N_-\cr
r_-&=Mx\qquad\qquad N_-={Nx/2}\cr
x&={P^2+Q^2\over M^2+N^2}\qquad\ y={MQ+NP\over M^2+N^2} \ .\cr}}
Here, $M,$ $N$, $2\sqrt{2}P$ and $
2\sqrt{2}Q$, and $a$ are respectively the mass, NUT parameter,
magnetic and electric
charges, and the angular momentum per unit mass.

\def\ie{{\it i.e.,}\ }
\def\cam{{\cal M}}
\def\cak{{\cal K}}
\def\ssc{\scriptscriptstyle}
The dilaton and the scalar axion are\galtsov:
\eqn\thescalars{\eqalign{
e^{\Phi}&={\W/\S}\cr
\Ax&={1\over\W}\left[(2Py-Nx)r-QPx+(Mx-2Qy)\d\right]\cr}}
where we have chosen to set $\Phi\to0$ and $\Ax\to0$ in the
asymptotically flat region, \ie $r\to\infty$.
The time component of the gauge field is given explicitly
as\galtsov:
\eqn\potential{
A_t={2\sqrt{2}\over\S}\left[Q(r-r_-)+P\td\,\right]\ .}
Ref.\galtsov\ only presents an implicit definition for the
the spatial components of the gauge potential in terms
of a `magnetic' potential
\eqn\magpot{u={\sqrt{2}\over\S}\left[P(r-r_-)-Q\td\,\right]\ .}
$A_\phi$ is then determined through
\eqn\magfield{\eqalign{
F_{r\phi}&={\Omega\sin\theta\over\Delta-a^2\sin^2\theta}
\left(2\partial_\theta u+\rho\partial_\theta A_t
\right)-\omega\partial_r A_t\cr
F_{\theta\phi}&=-{\Omega\Delta\sin\theta\over\Delta-a^2\sin^2\theta}
\left(2\partial_\theta u+\rho\partial_\theta A_t
\right)-\omega\partial_\theta A_t\ .\cr}}

\vfill\eject

\subsection{The Extremal Limit}
We can describe the extremal solutions in terms
of four distinct regions\ref\gs{S. B. Giddings and A. Strominger,
{\sl Phys. Rev.} {\bf D46} (1992)  627, hep-th/9202004.}:
First, there is the near
neighbourhood of the horizon.
This region connects onto a `throat' region in which the
geometry is essentially constant.
This throat eventually widens out
at the `mouth' region, and finally connects onto the asymptotically
flat region. See figure \fig\theonlyfigure{A schematic depiction of the
three regions of the extremal geometry of the dyon. The radial coordinate
$\sigma$ runs along the throat from $\sigma=0$ (the horizon) in region (a). It
is possible to continue behind the horizon to reveal a singularity. A circle
here represents the remaining coordinates. $\Lambda$ represents the
length of the throat region, which diverges logarithmically at extremality.
See the text for further explanation.}:

\epsfxsize=5.0in\epsfbox{throat.eps}

The length  of the throat region $\Lambda$ %r moved $\Lambda$
diverges logarithmically as one approaches the
extremal limit. In examining the extremal solution, there are
several  ways to approach this limit leading to
three regions of the geometry %r extra words for figure
(illustrated in the figure): (a) the `horizon $+$ throat' solution,
which is approached by holding the horizon radius
fixed and letting
the mouth and asymptotically flat regions move off to infinity;
(b) the  throat solution,
which is derived by letting both the horizon and the
asymptotically flat region tend to infinity; and %r reorder b and c
(c) the throat $+$ asymptotically flat solution, which is found
by fixing the mouth and asymptotically flat region while %r and
taking the horizon off to infinity.
These different solutions are derived by carefully tuning
the parameters of the full solution\gs.
We expect that the heterotic coset solution \thesolution\
describes the `horizon $+$ throat' limit of the extremal
version of the above low energy solution.

\subsubsection{The `Horizon $+$ Throat' Region}
Following ref.\gs\ to uncover this geometry,
we make a coordinate transformation $r=r_{\ssc H}+\g f(\s)$
where $r_{\ssc H}$ is the position of the horizon, and
$\g$ is a small `scaling' parameter. This focuses our analysis
on the neighbourhood of the horizon.
The scaling parameter $\g$ will at the same time control
how the parameters in the solution deviate from their
extremal values as we approach extremality.

In the full solution the `horizons' occur at $G^{rr}=0$, \ie
\eqn\horizon{
0=\D=r^2-2M(1+{x\over2})\,r+2xM^2+a^2-N^2(1-{x\over2})^2\ .}
This gives the `horizon' positions as:
\eqn\horizons{r_{\ssc H\pm}
=M\left(1+{x\over2}\right)\pm\sqrt{(M^2+N^2)\left(1-{x\over2}\right)^2-a^2}
\ . }
At extremality these positions coincide, \ie the second term above
\eqn\distance{ D=\sqrt{(M^2+N^2)\left(1-{x\over2}\right)^2-a^2}} vanishes.
Hence in our `scaling' limit, we will have $D\propto\g$.

Also of interest in the rotating solutions is the ergosurface (at which
time translations are null) which is given by $G_{tt}=0$:
\eqn\ergosurface{
0=\D-a^2\sin^2\th
=r^2-2M(1+{x\over2})\,r+2xM^2+a^2\cos^2\th-N^2(1-{x\over2})^2\ .}
Hence the position of the ergosurface is
\eqn\ergorad{\eqalign{
r_{\ssc E}=&M\left(1+{x\over2}\right)+
\sqrt{(M^2+N^2)\left(1-{x\over2}\right)^2-a^2\cos^2\th}\cr
=&r_{\ssc H+}+\sqrt{(M^2+N^2)\left(1-{x\over2}\right)^2
-a^2\cos^2\th}-\sqrt{(M^2+N^2)\left(1-{x\over2}\right)^2
-a^2}.\cr} }
Now examining the heterotic coset solution \thesolution, we see that the
ergosurface occurs entirely at finite $\sigma$, suggesting that
we should choose $r_{\ssc E}-r_{\ssc H+}\propto\g$ as well.
Thus from eq.\ergorad, %r remind the reader to look at the equation
%r we can only introduce
we see only an infinitesimal amount of
angular momentum can be introduced %r
into the scaled `horizon $+$ throat' solution, \ie
$a\propto\g$ as well.

With $a$ simply set to zero, we recover the familiar extremal solution\us\them\
with $x=(Q^2+P^2)/(M^2+N^2)=2$. In preparation for our
extremal  limit, we set
\eqn\near{
\eqalign{
a&=\g\a\cr
x&=2-\g{2\over \sqrt{M^2+N^2}}\cr
}}
which gives
\eqn\then{D=\gamma\sqrt{(1-\alpha^2)\ .}}
As described above, the small parameter
$\gamma$ controls how close we are to the extremal limit.
For our radial coordinate, we choose
\eqn\scaling{r=r_{\ssc H+}+\g f(\sigma) }
where as above $r_{\ssc H+}=M\left(1+{x\over2}\right)+D$ is the position
of the event horizon, and $\sigma$ will be our new `scaled' \co\ in the
$\g\to0$ limit.

The final parameters, which we should consider `scaling',
are the electric and magnetic charges. At $\g=0$, we choose charges:
$Q=Q_o$ and $P=P_o$ such that
\eqn\chargerelation{
Q_o^2+P_o^2=2(M^2+N^2)
}
since $x=2$.
For non--vanishing $\g$, $x$ is not precisely 2 as given in
eq.~\near, and so we choose
\eqn\nearcharge{
\eqalign{
Q&=Q_o-(1-\b)\g{Q_o^2+P_o^2\over2Q_o\sqrt{M^2+N^2}}\cr
P&=P_o-\b\g{Q_o^2+P_o^2\over2P_o\sqrt{M^2+N^2}}.\cr
}}
It will turn out that the parameter $\b$ does not enter into
our final solution.
It remains to make precise a choice for
$Q_o$ and $P_o$, but in fact this choice is only restricted in
a minimal way. Since we anticipate that a
throat geometry will arise in the $\g\to0$ limit,
$G_{\theta\theta}=\W$ should become a (finite) constant. Inserting
our scaling ansatz\"e for the various parameters, we find
\eqn\nearthroat{
\W=\W_o=2{(MP_o-NQ_o)^2\over M^2+N^2}+O(\g)
}
where we have used $Q_o^2+P_o^2=2(M^2+N^2)$ to simplify the
final expression. Hence we get the desired constant behavior
unless the numerator vanishes. One can show this only occurs for
$(Q_o,P_o)=(\sqrt{2}M,\sqrt{2}N)$ or $(-\sqrt{2}M,-\sqrt{2}N)$.
So we may choose any charges satisfying equation \chargerelation\
except within $O(\g)$ of these special values.\foot{In fact, one may
produce a slightly more subtle scaled solution even for these disallowed
charges, but they do not contain a constant throat geometry. They
correspond to the $S$-dual solutions to the conformal field
theory solutions considered here.}

Now it only remains to find an appropriate function $f(\sigma)$ in
the radial coordinate. For simplicity, we will require that $f(\sigma=0)=0$
so that the horizon corresponds to $\sigma=0$.
Examining equation \thesolution\ shows that
the line element has $dS^2\simeq k (d\sigma^2+d\theta^2)$.
Using $G_{\theta\theta}=\Omega_o$,
$f(\sigma)$ is uniquely determined to be
$f(\s)=\sqrt{1-\alpha^2}(\cosh\s-1)$ by the equation
 ${\W\over\D}dr^2= \Omega_o d\s^2$.

To determine the rest of background fields in the throat limit,
we substitute
\eqn\thelot{\eqalign{
r&=M(1+{x\over2})+\gamma\sqrt{(1-\alpha^2)}\cosh\sigma\cr
a&=\gamma\alpha\cr
{\rm and }\qquad\qquad x&=2-2{\gamma\over\sqrt{M^2+N^2}}
}}
into eqs.\galtsovmetric, \now, \thescalars, \potential,
\magpot\ and \magfield, and take the limit $\g\to0$.

Taking $G_{\tt\tt}$ for example, we find that in this limit, after comparison
with \thesolution, leads to
\eqn\gttthree{G_{\tt\tt}=
{\W_o\over4M^2}{\tau^2\sin^2\theta-\sinh^2\sigma
\over(\lambda\tau\cos\theta+\cosh\sigma
+\delta)^2},}
if we make the identification
\eqn\identify{\eqalign{
\tau&={\alpha\over\sqrt{1-\alpha^2}}\cr
\delta&={\sqrt{1+{N^2\over M^2}}\over\sqrt{1-\alpha^2}}\cr
{\rm and}\ \ \lambda&=-{N\over M}\ \ .\cr
}}
Finally we perform a rescaling $\tt=2M\,t$ in order to match the
overall factor of $\W_o$ of $G_{\s\s}$ and $G_{\theta\theta}$.
In similar fashion the metric components
$G_{\phi\phi}$ and $G_{\phi t}$ arise  as identical to those
given in \thesolution.
Note that the parameters satisfy a
relation
\eqn\relation{(1+\tau^2)(1+\lambda^2)=\delta^2\ .}
Examining the low energy limit of the anomaly
equations \canceltwo\ it is easy to see that this same
relation
arises upon the algebraic elimination of the charges $Q_A$
and $Q_B$.

Also, the dilaton \thescalars\ of ref.\galtsov\ becomes that in \thesolution\
in the extremal limit, after the (now familiar\gps\us\gs) consistent
absorption of an
infinite additive constant into $\Phi_0$ --- \ie $\Phi_0=- %r sign
\log[2\g\sqrt{1-\alpha^2}/\W_o]$. Thus the dilaton is tuned to make it
finite in this region of interest.
Similarly from \thescalars, we recover the axion field
in the extremal limit as:
\eqn\hisaxion{\Ax-\Ax_0 =e^{-\Phi_0}(\lambda\cosh\sigma+\tau\cos\theta),}
which is the same as \ouraxion\ (and $\Phi_0$ is the constant given above).
Here  we have additionally
shifted an infinite additive constant into $\rho_0$ to make
$\rho$ finite in this region of interest. As the axion is only defined up to a
constant in \defax, this is also a consistent operation.

Turning our attention to the associated gauge fields of
ref.\galtsov, we find that in eq.\potential\ our limit yields:
\eqn\afield{A_t={2\sqrt{2}}(P_o+Q_o\lambda){\tau\cos\theta-
{\lambda\delta\over1+\lambda^2}\over
\cosh\sigma+\delta-\lambda\tau\cos\theta}
+{2\sqrt{2}Q_o}\ .
}
Note that the subscript index here is $t$ rather than $\tt$.
Here, we need first to make a trivial gauge shift to remove the constant term
in this expression. Then,
the  result is precisely our gauge field from \thesolution\
if we identify
$Q_B=(P_o+Q_o\lambda)$, and use $Q_A={\lambda\delta\over1+
\lambda^2}Q_B$ from eq.\canceltwo.
With the same identifications, one finds that scaling \magpot\
and \magfield\ produces the $A_\phi$ that appears
in equation \thesolution. Also note that this identification of the
charges along with eq.\canceltwo\ yields $\Omega_o=k$, as is required
for the overall factor in the metric.

\subsubsection{The Other Regions}
The asymptotically flat region is obtained  by taking again the
$\g\to0$ limit for the parameters as in \near\ and \nearcharge, but with
$r=r_{\ssc H+}+y$ with $y$ fixed and
large. Thereby the horizon recedes an infinite distance from
the asymptotic region. The metric and accompanying  fields become:
\eqn\flatass{\eqalign{
dS^2&=
F(y)\left[-\left(1+{2M\over y}\right)^{-2}(d\tt-2N\cos\theta  d\phi)^2+
dy^2+y^2(d\theta^2+\sin^2\theta d\phi^2)\right]\cr
\Phi-\Phi_o&=\log\left({yF(y)\over (y+2M)}\right)\cr
\rho&={[2MP_0Q_0+N(P_0^2-Q_0^2)](y+2M)-P_0Q_0(M^2+N^2))\over y^2F(y)}\cr
A_\tt&={2\sqrt{2}Q_0\over y+2M}\cr
u&={2\sqrt{2}P_0\over y+2M}
,}}
where
\eqn\fy{F(y)=1+2{P_0(P_0M-Q_0N)\over (M^2+N^2)}{1\over y}
+2{(P_0M-Q_0N)^2\over (M^2+N^2)}{1\over y^2},}
which smoothly connects to the `other side' of the infinitely long
throat region, as can be seen by taking $y=2M\,\e{\sigma}$ for $\sigma$ large
and negative:
\eqn\purethroat{\eqalign{
dS^2&=\Omega_0\left( d\sigma^2-(dt-\lambda\cos\theta d\phi)^2+
d\theta^2+\sin^2\theta d\phi^2\right)\cr
\Phi-\hat{\Phi}_0&=-\sigma.}} Precisely this behaviour can be recovered in the
large $\sigma$ limit of the `horizon $+$ throat' geometry \thesolution.
(Recall %r (Note
that in the above $u$ is the magnetic potential %r field
defined in \galtsov\ from which $A_\phi$
%r the magnetic component of the gauge sector
may be derived via the appropriate limit of eq.\magfield.)

As stated before, it is an open problem as to how to explicitly construct the
conformal field theory description of the connection of
 the `horizon $+$ throat'
region (which we have successfully described as a CFT) to this `throat $+$
asymptotically flat' region.

\section{A Natural Generalisation}
Examining  the choice of gauge symmetries \gaugingstwo\ which
gave us our extremal
Kerr--Taub--NUT solution, the following highly symmetric pattern of
gaugings suggests itself as a generalisation:
\eqn\gaugingsthree{U(1)_A\times U(1)_B:\left\{\jump\eqalign{
g_1\to&\e{\epsilon_A\sigma_3/2}g_1
\e{(\delta\epsilon_A+\lambda\epsilon_B)\sigma_3/2}\cr
g_2\to&\e{i(\tau\epsilon_A+\eta\epsilon_B)
\sigma_3/2}g_2\e{i\epsilon_B\sigma_3/2}\ .}\right.}
Here, one might speculate about
what the spacetime interpretation of the parameter $\eta$ should
be. Clearly it produces a new axisymmetric coupling.
We carried out the procedure of
constructing the full conformal field theory, using the techniques
described in section~2 and then took the low energy limit in the manner
described in section~3.
We shall not repeat those steps here as they are essentially unchanged. However
note that in the Euler parameterisations \euler\ and \euleri\ the gauge
transformations of the fields are the same as in \transform\ with the
exception of $\phi$ which is now also translated under
the action of $U(1)_B$ %r old: addition of
\eqn\newact{\phi\to\phi+\tau\epsilon_A+\eta\epsilon_B\ .}
%r added \epsilon_A stuff above
%r old: for the action of $U(1)_B$.

Also the WZW gauging anomalies are modified to include contributions
$-k_2\eta^2$ for the $BB$ sector and $-k_2\eta\tau$ for the $AB$ (=$BA$)
sector. With the new action \newact\ of $U(1)_B$ on $\phi$, we might anticipate
some modification of the periodicity of $\phi$ in the final low--energy metric,
in a way analogous to the modification discussed in the case of the $\lambda$
coupling previously (see subsection 2.2).

\def\ins{(\eta\cos\theta+1)}
\def\hp{{\hat\phi}}
\def\tp{{\tilde{\phi}}}
\subsection{The Low Energy Limit in Three Gauges}
We display the solution in the three most transparent world sheet
gauges (referred to as (1), (2) and (3)) as it is instructive and shall
facilitate further discussion.

(1) Using the same worldsheet gauge as previously ($t_L=0=\psi$), we get:
\eqn\thesolutionthree{\eqalign{dS^2&=k\Biggl[d\sigma^2+d\theta^2
-\left({\sinh\sigma(\ins dt-\lambda\cos\theta d\phi)\over
\ins(\cosh\sigma+\delta)
-\lambda\tau\cos\theta}\right)^2
\cr &\hskip1cm
+\left({\sin\theta(\tau dt-(\cosh\sigma+\delta)d\phi)
\over\ins(\cosh\sigma+\delta)
-\lambda\tau\cos\theta}\right)^2\Biggr] \cr
\Phi&=-\log[\ins(\cosh\sigma+\delta)-\lambda\tau\cos\theta]+\Phi_0\cr
{B}_{t\phi}&=-k{\tau+\lambda\cos\theta\cosh\sigma\over\ins
(\cosh\sigma+\delta)-\lambda\tau\cos\theta}\cr
{A}_t&=-{2\sqrt{2}(Q_A(\eta\cos\theta+1)
-\tau Q_B\cos\theta)\over\ins(\cosh\sigma+\delta)-\lambda\tau\cos\theta }\cr
{ A}_\phi&=-{2\sqrt{2}\cos\theta(Q_B(\cosh\sigma+\delta)-\lambda Q_A)\over\ins
(\cosh\sigma+\delta)-\lambda\tau\cos\theta}\ ,}}
showing the $\eta$--generalisation of \thesolution.

(2) The solution in the $t_L=0,\psi=\mp\phi$ gauge is:

\def\AM{A^M(\theta)}
\eqn\thesolutionfour{\eqalign{
d{\hat S}^2&=k\Biggl[d\sigma^2+d\theta^2
-\left({\sinh\sigma (\ins dt+2\lambda\AM d\hp)
\over\ins(\cosh\sigma+\delta)
-\lambda\tau\cos\theta}\right)^2
 +\cr
&\hskip1cm+\left({\sin\theta(\tau dt\mp((\eta\pm1)(\cosh\sigma+\delta)
-\lambda\tau)d\hp)
 \over\ins(\cosh\sigma+\delta)
-\lambda\tau\cos\theta}\right)^2\Biggr] \cr
\Phi&=-\log[\ins(\cosh\sigma+\delta)-\lambda\tau\cos\theta]+\Phi_0\cr
{\hat B}_{t\hp}&=-2k{\AM(\pm\tau-\lambda\cosh\sigma)\over\ins
(\cosh\sigma+\delta)-\lambda\tau\cos\theta}\cr
{\hat A}_t&=-{2\sqrt{2}(Q_A(\eta\cos\theta+1)
-\tau Q_B\cos\theta)\over\ins(\cosh\sigma+\delta)-\lambda\tau\cos\theta }\cr
{\hat A}_\hp&={4\sqrt{2}\AM[Q_B(\cosh\sigma+\delta)-\lambda Q_A]\over\ins
(\cosh\sigma+\delta)-\lambda\tau\cos\theta}\ ,}}
where, as usual the monople field is $$\AM\equiv {\pm1-\cos\theta\over2}.$$

(3) There is also the gauge choice $t_R=t_L=0$. This one is not valid at
$\lambda=0$, but is useful as it makes the $S^3$ topology manifest:

\eqn\thesolutionfive{\eqalign{
d{\tilde S}^2&=k\Biggl[d\sigma^2+d\theta^2
-\left({\lambda\sinh\sigma(d\psi+\cos\theta d\tp)
\over\ins
(\cosh\sigma+\delta)-\lambda\tau\cos\theta}\right)^2
\cr &\hskip1cm
+\left({\sin\theta\left\{(\lambda\tau-\eta(\cosh\sigma+\delta))
d\psi+(\cosh\sigma+\delta)d\tp\right\}
\over\ins(\cosh\sigma+\delta)
-\lambda\tau\cos\theta}\right)^2\Biggr]
\cr
\Phi&=-\log[\ins(\cosh\sigma+\delta)-\lambda\tau\cos\theta]+\Phi_0\cr
{\tilde B}_{\psi\tp}&=k
{(\cos\theta+\eta)(\cosh\sigma+\delta)-\lambda\tau
\over\ins(\cosh\sigma+\delta)-\lambda\tau\cos\theta}\cr
{\tilde A}_\psi&={2\sqrt{2}[Q_B(\cosh\sigma+\delta)-\lambda Q_A]\over\ins
(\cosh\sigma+\delta)-\lambda\tau\cos\theta}\cr
{\tilde A}_\tp&=-{2\sqrt{2}\cos\theta[Q_B(\cosh\sigma+\delta)
-\lambda Q_A]\over\ins
(\cosh\sigma+\delta)-\lambda\tau\cos\theta}\ .}}

For all forms of the solution, the  low--energy form of the anomaly equations
relates the parameters in the fields:
\eqn\cancelagain{\eqalign{
{k\over2}={Q_A^2\over\delta^2-1-\tau^2}=
{Q_B^2\over1+\lambda^2-\eta^2}={Q_AQ_B\over\delta\lambda-\eta\tau}.
}}
%r old: (As before, the $\lambda=0$ case needs to be supplemented with the
%$P_A,P_B$ charges for fully dyonic solutions.)
%r new:
Note that in this case, dyonic solutions are possible at $\lambda=0$
with a single $U(1)$ field.

\subsection{World Sheet and Spacetime Symmetries Revisited}
It is instructive to pause here to note how  the freedom
to change the  world sheet gauge slice implements spacetime symmetry
transformations. The sets
of
spacetime fields resulting from the three distinct
world sheet gauge choices above
 should be related to each  by spacetime \co\ transformations and gauge
transformations of the $U(1)$  gauge and antisymmetric tensor fields as
usual in heterotic string theory\ref\gsw{M. B. Green, J. Schwarz and E.
Witten, {\sl `Superstring Theory'}, Cambridge University Press, 1987.}:
\eqn\spacetimegauge{\eqalign{X^\mu\to &{X^\prime}^\mu(X)\cr
A\to & A+d\Lambda^{(0)}\cr
B\to & B+{1\over\sqrt2}A\, %r \wedge
d\Lambda^{(0)}+ d\Lambda^{(1)}, }}
where $\Lambda^{(0)}$ and $\Lambda^{(1)}$ are arbitrary zero-- and
%r new footnote
one--forms\foot{Under $U(1)$ gauge transformations, one usually
writes $\delta B={1\over\sqrt2}\Lambda^{(0)}dA$. The difference
with our variation in \spacetimegauge\ can be absorbed with
an additional antisymmetric tensor transformation, \ie
$\delta B={1\over\sqrt2}A\,d\Lambda^{(0)}+{1\over\sqrt2}d(\Lambda^{(0)}
A)$.}.

The $U(1)_B$  gauge transformations are given in \transform\ with \newact,
and the different world sheet gauges  are easily seen to be related to
one another. The processes of sections~2 and~3 ensure that the low energy
solutions above are related to each other via gauge and
coordinate  transformations of the form \spacetimegauge, which find their roots
in the worldsheet transformations \transform\ (with \newact). (Note that this
would not have worked without the crucial step of refermionisation mentioned in
section~3 and discussed in detail in ref.\cvj.)

The relation of solution (1) to solution (2) is the coordinate transform:
\eqn\onetwo{\eqalign{t\to&t\mp\lambda\hp\cr
\hp\to&{\phi\over(1\pm\eta)}},}
and the gauge transformations:
\eqn\gaugeonetwo{\eqalign{
{\hat{A}}=&\hat{A}_tdt+\hat{A}_\hp d\hp\to \hat{A}_tdt+(\hat{A}_\hp\mp
\lambda\hat{A}_t) d\hp\cr
=&{A}_tdt+({A}_\phi\pm{2\sqrt{2}Q_B\over(1\pm\eta)}) d\phi\cr
{\hat B}=&\hat{B}_{t\hp}dt\wedge d\hp\cr
=&\left[B_{t\phi}\pm {1\over(1\pm\eta)}\left( \lambda k+
{Q_B\over\sqrt2}A_t\right)  \right]dt\wedge d\phi.}}

The relation of gauge (3) to gauge (1) is the coordinate transformation:
\eqn\threetwo{\eqalign{
\phi=&\tp+{\eta\over\lambda}t\cr
t=&\lambda\psi},}
and the gauge transformations:
\eqn\gaugethreetwo{\eqalign{{{A}}=&{A}_tdt+{A}_\phi
d\phi\to (\lambda{A}_t+\eta A_\phi)d\psi
+{A}_\phi d\tp\cr
=&(\tilde{A}_\psi-2\sqrt{2}Q_B) d\psi+\tilde{A}_\tp d\tp\cr
{\tilde B}=&\tilde{B}_{\psi\tp}d\psi\wedge d\tp\cr
=&\left[\lambda B_{t\phi}+\eta k-
{Q_B\over\sqrt2}A_\phi  \right]d\psi\wedge d\tp.}}

\vfill\eject
\subsection{Spacetime Geometry} %r old: Interpretations}
For $\lambda\neq0$ it is most instructive to look at the solution in gauge (3).
There we see that constant $\sigma$ slices posses the topology of
$S^3$, inherited from the $SU(2)$ sector of the parent group of the coset.
%r old: The model's constant $\sigma$ metric is simply a `deformed'
%version of that for $S^3$.
%r new:
The metric on surfaces of constant $\sigma$ is that of a `deformed'
three--sphere, and in particular, there are no conical singularities.

For $\lambda=0$, where now $(\theta,\phi)$ parameterise an $S^2$,
there is the possibility that $\eta$ would
parameterise  conical singularities running along the $\theta=0$ or $\pi$
axes, by changing the periodicity of $\phi$.
Note that on the original three sphere
we have the following identifications (see \euler)
\eqn\idi{(\psi,\phi)\simeq(\psi,\phi+4\pi)\simeq(\psi-2\pi,\phi+2\pi)
\simeq(\psi
+2\pi,\phi+2\pi).}
Now the gauging imposes the extra identification
\eqn\extraid{
(\psi,\phi)\simeq(\psi+x,\phi+\eta x).}
Combining these we have
\eqn\combinedid{
(\psi,\phi)\simeq(\psi,\phi+2\pi(1+\eta))\simeq(\psi,\phi+2\pi(1-\eta))}
so if we gauge fix $\psi=0$, we have
\eqn\gaugedtozero{\phi=\phi+4\pi=\phi+2\pi(1-\eta)=\phi+2\pi(1+\eta).}
%r old:
%Therefore the field $\phi$ which appears in the final action in gauge
%(1) has fundamental period given by $2\pi\nu$
%where $\nu$ is the greatest common factor of the
%numbers\foot{This assumes that $\eta$ is a rational number.
%If it were not, the action of $U(1)_B$ on the
%compact $SU(2)$ sector would be ill--defined.}\ $\{2,1+\eta,1-\eta\}$.
%r new
In this case we see that $\eta$ must be a rational number,
otherwise the action of $U(1)_B$ on the compact $SU(2)$ sector
would be ill--defined. With rational $\eta$ then,
the field $\phi$ which appears in the final action in gauge (1) has
fundamental period given by $2\pi\nu$ where $\nu$ is the greatest
common factor of $\{2,1+\eta,1-\eta\}$. %r more new:
Hence this gauging makes orbifold identifications\ref\orbifold{L.
Dixon, J. Harvey, C. Vafa and E. Witten, {\sl Nucl. Phys.} {\bf B261}
(1985) 678; {\bf B274} (1986) 285.} on the angular two--spheres.
Now let us examine the solution in gauge (1) as in %r words,
eq.\thesolutionthree. Setting $\lambda=0$ and travelling in
 small loops about the $\theta=0,\pi$ axis for fixed values of the other
fields let us examine the form of the angular %r word
line element
%r replace d\varphi^2 by d\Omega^2
$d\Omega^2=d\theta^2+G_{\phi\phi}d\phi^2$ in the neighbourhood
of $\theta=0,\pi$.  We see that
$d\Omega^2=d\theta^2+\sin^2\!\theta\, d\phi^2/(1\pm\eta)^2$.
%r new:
The standard line element on $S^2$ is $d\Omega^2=d\theta^2+
\sin^2\!\theta\,d\hp^2$ with $\hp$ having period $2\pi$,
%r old: For $S^2$ we require that $d\Omega^2=d\theta^2+\sin^2\!\theta\,d\hp^2$
%with $\hp$ having period $2\pi$,
otherwise there is a  conical singularity due to the deficit
angle. Here we see that to get the standard form of the line element
we set $\hp=\phi/(1\pm\eta)$. %r but $\hp$ would have
%r the period of $\phi$  divided by $(1\pm\eta)$.
We have deduced the period of $\phi$ from the gauging to be $2\pi\nu$,
%r old:and therefore in general, $\hp$ does indeed have  period other
%rthan $2\pi$.
and hence rather than $2\pi$, $\hp$ has period $2\pi\nu/(1\pm\eta)$.
We therefore  have conical singularities on the axes. %rin the metric.
%r new orbifold stuff:
It is interesting that the orbifold singularities are different
on the $\theta=0$ and $\pi$ axes.
If the solution of this section (for $\lambda=0$)  was
 taken to be the metric (and associated
fields) of a
macroscopic extended massive
object %r leave up in text:\foot{A
(analogous to a cosmic string, perhaps) %r .},
aligned along the $\theta=0,\pi$ axis,
it would have a mass per unit length proportional to the
deficit angle $\epsilon$ which is a function of the parameter $\eta$.
We speculated that perhaps this solution was the
extremal limit of a generalisation of the  Kerr--Taub--NUT metric to include a
conical singularity as well as the stringy fields. Such solutions are known in
General Relativity (GR). For example, a limit may be taken of the known
solutions for collinear masses in GR (involving sending a neighbouring mass to
spatial infinity while sending its mass to infinity also) which produces a
conical singularity along the axis. Our  speculation was that perhaps upon
constructing the corresponding limit for two Kerr--Taub--NUT--type objects
aligned along a common  axis of rotation, the extremal limit might indeed yield
our solution of this section. So far, our initial attempts to demonstrate
this have not borne fruit.

 This  more general low--energy solution is interesting in its own
right however, regardless of whether it might be obtained as the
extremal limit of some as yet unknown
solution.

\section{Discussion}
In this paper we have explicitly constructed a family of
$(0,2)$ supersymmetric
conformal field theories as a heterotic coset models.
Our intention was to exploit
the geometric freedom inherent in the construction of such models to allow us
to use simple geometrical  intuition to introduce
a parameter which
would generate rotation in the low energy space time interpretation.
 Since this model was
constructed as a generalisation  of the Taub--NUT dyon theory in ref.\cvj,
it was natural to speculate that by slightly modifying the gauging
would  produce the dyonic Kerr--Taub--NUT solution.

\def\tchi{{\tilde{\chi}}}
\def\tOm{\widetilde{\Omega}_H}
We demonstrated this explicitly by taking the background fields
 of the stringy Kerr--Taub--NUT solution constructed in ref.\galtsov, and
showing that the `horizon $+$ throat' region of their  extremal limit
 coincides precisely with the
low--energy limit \thesolution\ of our conformal field theory.
  Despite the vanishing angular momentum as can be seen in \purethroat\ and
\flatass, the angular
velocity of the horizon remains finite at extremality.
One finds that the horizon generating Killing field for the low
energy solution is $\tchi^\mu\partial_\mu=\partial_\tt+\tOm\partial_\phi$
with
\eqn\lowvelocity{\tOm=
{2a\over\left(1-{x\over2}\right)(M^2+N^2)+M\,D}}
with $D$ defined in \distance.
In our scaling limit, this reduces to
\eqn\exlowvel{\tOm={1\over2M}{\tau\over1+\delta}={\Omega_H\over2M}\ .}
The difference between this extremal limit of $\tOm$, and $\Omega_H$
given in \angular, is precisely accounted for by the scaling of
the time coordinate, \ie $\tt=2M t$.

As mentioned in the introduction, the fact that an infinite family of solutions
labelled by $\tau$ may exist as the internal spacetime of the dyon,
smoothly connecting to a unique asymptotically flat extension with zero
angular momentum is interesting.
This is interesting in its own right, but is additionally so in the light of
the `remnant' proposals for a solution to the information puzzle\foot{The
authors are grateful to Petr Horava for drawing our attention to this.}.
Localised gravitational solutions which develop large internal spacetimes
(and reach   the  endpoint of their Hawking evaporation)  at extremality are of
great interest in these scenarios as not only do they form a stable
remnant, but they allow for the storage of any `lost' information (to the
asymptotic outside world) inside this `internal'
geometry\ref\remnants{Y. Aharonov, A. Casher and  S. Nussinov,
{\sl Phys. Lett.} {\bf B191} (1987) 51\semi
A. Casher and  F. Englert, {\sl Class. Quant. Grav.} {\bf 9}  (1992) 2231\semi
S. B. Giddings, {\sl Phys. Rev.}  {\bf D46} (1992) 1347, hep-th/9203059\semi
T. Banks, M. O'Loughlin and A.  Strominger, {\sl Phys. Rev.}  {\bf D47}
(1993) 4476
hep-th/9211030\semi
T. Banks and  M. O'Loughlin, {\sl Phys. Rev.}  {\bf D47} (1993) 540,
 hep-th/9206055\semi
A. Strominger and S. Trivedi, {\sl Phys. Rev.}  {\bf D48} (1993) 5778,
 hep-th/9302080\semi
S. B. Giddings, {\sl Phys. Rev.}  {\bf D49} (1994) 947, hep-th/9304027\semi
D.A. Lowe and  M. O'Loughlin, {\sl Phys. Rev.}  {\bf D48} (1993) 3735,
 hep-th/9305125\semi
J. D. Bekenstein, {\sl Phys. Rev.}  {\bf D49} (1994) 1912, gr-qc/9307035\semi
S. B. Giddings, {\sl Phys. Rev.}  {\bf D49} (1994) 4078,  hep-th/9310101\semi
P. Yi,  {\sl Phys. Rev.}  {\bf D49} (1994) 5295, hep-th/9312021\semi
L. Susskind, {\sl `Trouble for Remnants'}, Stanford Preprint hep-th/9501106.}.
In particular, a
remnant should have  an infinite number of internal states, degenerate from the
point of view of the outside world\foot{The additional fact that this internal
world is of infinite volume is regarded as a possible bonus
also, due to the fact that it  may suppress infinite pair production of such
remnants\ref\pairs{T. Banks, M. O'Loughlin, and A. Strominger,
{\sl Phys. Rev.}  {\bf D47} (1993)   4476, hep-th/9211030\semi
F. Dowker, J. P, Gauntlett, D. Kastor and J. Traschen, {\sl Phys. Rev.}  {\bf
D49}
(1994) 2909, hep-th/9309075\semi
F. Dowker, J. P, Gauntlett, S. B. Giddings and G. Horowitz, {\sl Phys. Rev.}
{\bf D50} (1994) 2662, hep-th/9312172}.}. In the present case we have such a
localised solution. At the bottom of its internal world we have a theory which
can be labelled by the parameter $\tau$, which can take arbitrary
%r ?rational? it doesn't have to be rational or am i forgetting something?
values, as allowed by the CFT. Upon traversing the throat and smoothly
connecting onto the asymptotic outside world, all reference to this `internal'
parameter is lost. Therefore we have (at least) one means of labeling the
infinitely degenerate internal state of this object. Note that this solution,
although in general dyonic, also
has neutral counterparts  with non--zero $\tau$
which thus serve as a generalisation of the `neutral remnant' solution of
ref.\gps. (See ref.\hetcosone\ for the heterotic coset model description of the
neutral solution of  ref.\gps.)
Of course, whether or not remnant proposals
are in general valid as solutions to the  information puzzle is a question
under intense
scrutiny.

Whether or not this picture survives
in the full classical string theory is an important question. As we only have
the CFT description of the region deep down the throat, we cannot rule out the
possibility that $\tau$ disappears completely from the external region only at
leading order in $\alpha^\prime$. It is conceivable that the complete classical
solution (which we will have when we learn how to connect the CFT
presented here to CFT's for the external spacetime) gives a unique continuation
from the  external spacetime to the internal sector at the level of a CFT.
In this case $\tau$ would survive the traversal of the infinite throat to the
outside region. It must be noted however that $\alpha^\prime$ corrections
cannot contribute to the leading order asymptotic form of the fields and
metric, as the corresponding higher derivative contributions in
the equations of motion %r old:as they are curvature effects which
are negligible in the asymptotic region.
Therefore, the conclusion made above that the angular momentum is zero will
be unaffected by whatever results await to be discovered in the CFT connection
to the outside region.

Note also that the parameter $\eta$  in the solution generalising  the rotating
dyon CFT which  we presented in section~5 will (in contrast to $\tau$) remain
in the metric all the way up the throat and also in any asymptotically flat
extension to the geometry, even at leading order, as can be seen from the large
$\sigma$ limit of any of the (equivalent) metrics in section~5. It therefore
is not to be considered as a `remnant'--like parameter in the sense suggested
by $\tau$.

Another interesting observation is that using the Killing field
$\tchi^\mu\partial_\mu=\partial_\tt+\tOm\partial_\phi$, the surface
gravity of the horizon becomes: $\tilde{\kappa}=\kappa/(2M)=
[2M(1+\delta)]^{-1}$. (Recall the surface gravity $\kappa$ was calculated in
eq.\surftwo. Also we assume $\lambda=0$ in the following.)
 Since the time
component of this Killing vector is normalized with $G_{\tt\tt}
\to-1$ in the asymptotically flat region, one may identify
$\tilde{\kappa}/(2\pi)$ as the Hawking temperature of the
horizon\radi. Note that examining the corresponding metric \flatass\
in the asymptotically flat region would have given a vanishing
surface gravity and Hawking temperature. This may be a more
appropriate description of the effective Hawking temperature,
since we expect that no Hawking radiation escapes to the
asymptotic region because of the infinite throat or the large
effective barrier near the horizon\chargedtwo\gs.

In general, we are encouraged that heterotic coset
constructions seem to be a powerful technique to obtain
the exact --- in the sense of having a
complete conformal field theory --- classical solutions to
many new interesting and important low energy backgrounds.
One might then
begin to examine other parent groups and gaugings, with  accompanying heterotic
arrangements of fermions.
One particularly interesting avenue of research would
be the {\it explicit}
 construction of conformal field theories corresponding to gauge and
gravitational instantons and
related objects\foot{See for example
ref.\edADHM\ for recent progress in such
constructions.}\ in all regions of their
geometry.

Once we have {\it some} conformal field theory and we have satisfied ourselves
that the low energy physics is interesting,
we must not forget that we should study its content, \ie\ spectrum,
correlation functions, moduli, etcetera,
to discover the stringy data which we wish
to learn about.
Such a program of study for heterotic coset models is currently in
progress\ref\hetcostwo{P. Berglund, C. V. Johnson, S.
Kachru and P. Zaugg, {\sl
`Heterotic Coset Models II:~The Spectrum of $(0,2)$ Conformal Field Theories'},
in preparation.}.

\acknowledgements

The authors would like to thank Petr Horava, Gary Horowitz
 and Amanda Peet for interesting
conversations. CVJ would like to thank
those at the Physics Department at McGill University (where some of this work
was carried out) for their kind hospitality in August 1994.
 CVJ is supported by a EPSRC (UK) postdoctoral
fellowship. RCM is supported by NSERC of Canada, and Fonds FCAR du
Qu\'ebec.

\listfigs
\medskip
\epsfxsize=5.0in\epsfbox{throat.eps}
\vfill\eject
\listrefs
\bye